\documentclass{aa}
\usepackage[varg]{txfonts}
\usepackage{braket}     
\usepackage{wrapfig}
\usepackage{gensymb}
\usepackage{hyperref}
\usepackage{booktabs}
\usepackage{color}
 
\usepackage{amstext}

\begin{document}

\title{Turbulent and wind-driven accretion in dwarf novae threaded by a large-scale magnetic field}
 \author{N. Scepi\inst{1},  G. Lesur\inst{1}, G. Dubus\inst{1} \and M. Flock\inst{2}}
 \authorrunning{N. Scepi et al.}
   \institute{Univ. Grenoble Alpes, CNRS, Institut de Plan\'etologie et d'Astrophysique de Grenoble (IPAG), F-38000, Grenoble, France
 \and
 Max Planck Institute for Astronomy, K\"onigstuhl 17, D-69117 Heidelberg, Germany}

\abstract{Dwarf novae (DNe) are accreting white dwarfs that show eruptions caused by a thermal-viscous instability in the accretion disk. The outburst timescales constrain $\alpha$, the ratio of the viscous stress to the thermal pressure, which phenomenologically connects to the mechanism of angular momentum transport. The eruptive state has $\alpha\approx0.1$ while the quiescent state has $\alpha\approx0.03$. Turbulent transport that is due to the magneto-rotational instability (MRI) is generally considered to be the source of angular momentum transport in DNe. The presence of a large-scale poloidal field threading the disk is known to enhance MRI-driven transport. Here, we perform 3D local magnetohydrodynamic (MHD) shearing-box simulations including vertical stratification, radiative transfer, and a net constant vertical magnetic flux to investigate how transport changes between the outburst and quiescent states of DNe. We find that a net vertical constant magnetic field, as could be provided by the white dwarf or by its stellar companion, provides a higher $\alpha$ in quiescence than in outburst, in opposition to what is expected. Including resistivity quenches MRI turbulence in quiescence, suppressing transport, unless the magnetic field is high enough, which again leads to $\alpha\approx0.1$. A major difference between simulations with a net poloidal flux and simulations without a net flux is that angular momentum transport in the former is shared between turbulent radial transport and wind-driven vertical transport. We find that wind-driven transport dominates in quiescence even for moderately low magnetic fields $\sim 1$ G. This can have a great impact on observational signatures since wind-driven transport does not heat the disk. Furthermore, wind transport cannot be reduced to an $\alpha$ prescription. We provide fits to the dependence of $\alpha$ with $\beta$, the ratio of thermal to magnetic pressure, and $T_{\rm eff}$, the effective temperature of the disk, as well as a prescription for the wind torque as a function of $\beta$ that is in agreement with both local and global simulations. We conclude that the evolution of the thermal-viscous instability, and its consequences on the outburst cycles of CVs, needs to be thoroughly revised to take into account that most of the accretion energy may be carried away by a wind instead of being locally dissipated.}


\maketitle

\section{Introduction}
The missing link between accretion theory and observations is the mechanism transporting angular momentum. Turbulent transport is one candidate and is often parameterized by the dimensionless parameter $\alpha$, ratio of the fluid stress to the local thermal pressure \citep{shakura}. Theory of thin viscous $\alpha$-disks has been widely applied to various systems such as protoplanetary disks (\citealt{sano2000}, \citealt{papaloizou1999}), soft X-ray transients (SXT) (\citealt{king1998}, \citealt{tetarenko2018}), dwarf novae (DNe) \citep{smak1984}, or even active galactic nuclei (AGN) \citep{starling2004}; see, however, \cite{hameury2009}). However, in soft X-ray transient sources and dwarf novae, which are the focus of this paper, the best estimators of $\alpha$ are often found because their nature is time dependent \citep{king2007}.

DNe are compact binary systems in which matter is transferred by Roche-lobe overflow from a solar-type star to a white dwarf. DNe have been observed over decades, and their light curves show periodic outbursts with amplitudes of 2-3 magnitudes \citep{Warner}. Their eruptive behavior is well explained by the disk instability model (DIM; \citealt{lasota2001}, \citealt{dubus2018}), in which a thermal instability caused by the ionization of hydrogen is responsible for a hysteresis cycle. In this system, the hot or cold state has a higher or lower accretion rate than the mass accretion rate coming from the companion, which causes the disk to fill up or empty. This leads to an hysteresis cycle in luminosity. The timescales involved are related to the ability for the disk to efficiently extract its angular momentum, providing an observational handle on $\alpha$. Outburst-decay timescales imply $\alpha\approx0.1$ in the hot state \citep{kotko2012}, and recurrence timescales gives $\alpha\approx0.01$ in the cold state (\citealt{cannizzo1988}, \citealt{cannizzo2012}). It is unclear why this change in $\alpha$ occurs and if it is related to the properties of turbulence. Nevertheless, we can use these observational constraints as a benchmark to distinguish between different transport mechanisms for DNe.

Consensus appears to be reached that the magneto-rotational instability (MRI; \citealt{balbus1991}) is the main driver of turbulent motions as it requires only a subthermal magnetic field threading the disk and a radially decreasing angular velocity; these are quasi-omnipresent conditions in astrophysical disks. However, magneto-hydrodynamical processes require coupling between gas and magnetic field, and this can be problematic in cold regions of protoplanetary disks \citep{gammie1996}, but also for dwarf novae in quiescence (\citealt{gammie1998}; \citealt{scepi2018} hereafter S18).

Given the complexity of MRI saturation, numerical simulations have been extensively used to study the characteristics of the resulting turbulence and transport. The properties of MRI-driven turbulence are known to depend on the magnetic configuration, especially the amplitude of the local net magnetic field \citep{Hawley1995}, which is largely unconstrained in DNe disks. A self-sustaining version of the MRI turbulence also exists in the absence of an externally imposed magnetic field, known as zero net-flux configuration (ZNF), and it is known to be independent of the initial magnetic seed \citep{hawley1996}. Hence, ZNF simulations provide a minimum level of angular momentum transport by the MRI that does not require a large-scale magnetic field. We add the caveat that the convergence of ZNF simulations is still debated (see \cite{ryan2017} for more information on the subject).

Isothermal simulations of MRI with ZNF are known to give a universal value of $\alpha\approx0.01$ (\citealt{hawley1996}; \citealt{simon2012}) that is characteristic of the quiescent state of DNe. Using an approximate function of cooling, \cite{latter2012} were able to retrieve thermal equilibrium curves of DNe from ZNF MRI simulations, showing the two expected stable branches, the hot and the cold state, which correspond to the eruptive and the quiescent states, respectively. However, their simulations exhibit a value of $\alpha\approx0.01$ regardless of the branch. By including vertical stratification, a realistic equation of state and radiative transfer in the regime of DNe, \cite{Hirose2014} found for the first time that ZNF MRI can lead to $\alpha\approx0.1$ in the low-density part of the hot branch. The actual interpretation is that there exists a non-linear coupling with convection which enhances $\alpha$ in this regime. This result was confirmed in S18 with a different code and in \cite{coleman2018} in a different regime of opacities. However, this enhanced $\alpha$ at the tip of the hot branch fails to reproduce the light curves of DNe by inducing reflares, which is inconsistent with observations \citep{coleman2016}. Moreover, S18 found that in ZNF the cold branch was too resistive to sustain magnetic turbulence, leading to $\alpha\approx 0$ in quiescence, as suggested by \cite{gammie1998}.

Hence, the simple application of ZNF MRI to the case of DNe leads to inconsistencies with the DIM model and the observations. ZNF is a common assumption since the magnetic configuration is unknown in DNe. Moreover, large-scale magnetic fields leads to higher Alfv\'en speeds in the atmosphere and thus are more computationally demanding. They also require a robust solver when the magnetization is high. It is well known, however, that $\alpha$ depends on the ratio of the thermal to magnetic pressure, $\beta$, \citep{hawley1996}, which can lead to $\alpha\gtrsim0.1$ if the disk is threaded by a strong enough large-scale magnetic field. Additionally, it is expected that a large magnetic field will sustain MRI turbulence to lower ionization levels than in ZNF \citep{fleming2000}. This readmits MRI as a possible candidate on the cold branch. The purpose of this paper is twofold: to study the local thermal equilibrium of a disk in realistic conditions of DNe threaded by a large-scale magnetic field, and to determine whether this field could resolve the discrepancies between MRI transport and observations.

\section{Methods}

We adopted the local shearing-box approximation \citep{Hawley1995} to simulate a vertically stratified patch of an accretion disk located at a distance  $R_0 = 1.315\times10^{10}\:\mathrm{cm}$ from a $0.6\,M_\odot$ white dwarf, giving an angular velocity $\Omega(R_0)=5.931\times10^{-3}\:\mathrm{s^{-1}}$ to facilitate comparisons with \cite{Hirose2014}. The simulations include radiative transport in the flux-limited diffusion approximation and  thermodynamic quantities appropriate to the temperature and density regime sampled by DNe. 

\subsection{Basic equations}
Curvature terms are not taken into account in the shearing-box approximation, and the differential Keplerian velocity is modeled as a linear shear flow $v_y^0=-(3/2)\Omega x$, where $x$, $y$, and $z$ correspond to the radial, azimuthal, and vertical directions, respectively. The set of equations in the corotating frame is
\begin{gather}
\frac{\partial\rho}{\partial t}+\pmb{\nabla}\cdot(\rho\pmb{v}) = 0,  \\
\begin{split}\label{eq:angular}
\rho\frac{\partial\pmb{v}}{\partial t}+(\rho\pmb{v}\cdot\pmb{\nabla})\pmb{v} = -\pmb{\nabla}\left(P+\frac{B^2}{8\pi}\right)+\left(\frac{\textbf{B}}{4\pi}\cdot\pmb{\nabla}\right)\textbf{B} +\rho\big(-2\Omega\pmb{\hat{z}}\times\pmb{v}\\ 
+ 3\Omega^2x\pmb{\hat{x}}-\Omega^2z\pmb{\hat{z}}\big),
\end{split}\\
\frac{\partial E}{\partial t}+\pmb{\nabla}\cdot[(E+P_t)\pmb{v}-(\pmb{v}\cdot\textbf{B})\textbf{B}] = -\rho\pmb{v}\cdot\pmb{\nabla}\Phi-\kappa_P\rho c(a_RT^4-E_R), \\
\frac{\partial\textbf{B}}{\partial t} = \pmb{\nabla}\times(\pmb{v}\times\textbf{B}-\frac{4\pi}{c}\eta\textbf{J}).
\end{gather}
The last three terms of Eq. (\ref{eq:angular}) represent the Coriolis force, the tidal force, and the vertical component of the gravitational force, respectively; $\pmb{\hat{x}}$ and $\pmb{\hat{z}}$ are the unit vectors in the $x$ and $z$ directions.

Radiative transfer is treated separately from the MHD step using an implicit time-stepping following the implementation of \cite{Flock2013}. In this step, we solve the coupled matter-radiation equations in the flux-limited diffusion approximation 
\begin{gather*}
\frac{\partial E_R}{\partial t}-\pmb{\nabla}\frac{c\lambda(R)}{\kappa_R\rho}\pmb{\nabla}E_R = \kappa_P\rho c(a_RT^4-E_R) \\
\frac{\partial \epsilon}{\partial t} = -\kappa_P\rho c(a_RT^4-E_R), 
\end{gather*}
where $\rho$ is the density, $\pmb{v}$ the fluid velocity vector, $P$ the thermal pressure, $\textbf{B}$ the magnetic field vector, $\Phi = \Omega^2(\mathrm{R_0})(-3x^2+z^2)/2$ is the gravitational potential in the corotating frame, $\eta$ the Ohmic resistivity, $\textbf{J}=(c/4\pi)\pmb{\nabla}\times\textbf{B}$ the current density vector, $E=\epsilon+0.5\rho\pmb{v}^2+\textbf{B}^2/8\pi$ the total energy, $\epsilon$ the internal energy, $P_t=P+\textbf{B}^2/8\pi$ the thermal pressure plus the magnetic pressure, $c$ the speed of light, $a_R=(4\sigma/c)$ the radiation constant with $\sigma$ the Stefan-Boltzmann constant, $T$ the temperature, $E_R$ the radiation density energy, $\kappa_P$ the Planck opacity, and $\kappa_R$ the Rosseland opacity. The radiative energy flux in the flux-diffusion approximation is $\textbf{F}_\mathrm{rad}= (c\lambda(R)/\kappa_R(T)\rho)\pmb{\nabla}E_R$. The flux limiter is defined as $\lambda(R)\equiv(2+R)/(6+3R+R^2)$ with $R\equiv |\nabla E|/(\kappa_R\rho E)$ \citep{turner2001}. We performed two simulations with the limiter from \cite{minerbo1978} and found no notable differences. We did not take into account radiation pressure as it is negligible for the temperatures reached by DNe; \citet{Hirose2014} found that it contributes 6\% of the gas+radiation pressure at most. 

To close our set of equations, we used the following equation of state (EOS) and internal energy function: 
\begin{gather*}
P = \frac{\rho}{\mu(\rho,T)}k_BT \\
\epsilon = \epsilon(\rho,T),
\end{gather*}
where $\mu$ is the mean molecular weight. To compute $\mu$ and $\epsilon$, we used precalculated tables and interpolated linearly between the table values. Tables were computed from the Saha equations assuming ionization equilibrium for the solar composition of \cite{grevesse1998} with a hydrogen abundance $X=0.7$ and a metallicity $Z=0.02$. The adiabatic index $\Gamma$, the entropy, and the thermal capacity $C_v$ were computed similarly. 

We used the opacity tables of \cite{Ferguson2005}, which cover the low-temperature region from $2.7 < \log(T) < 4.5$,  and OPAL \citep{Iglesias1996}, which covers the high-temperature region from $3.75 < \log(T) < 8.7$. We used a linear interpolation to connect the two  and extended the resulting table, where necessary, using a zero-gradient extrapolation.

\subsection{Boundary conditions}
We used shear-periodic boundary conditions in the $x$-direction and periodic boundary conditions in the $y$-direction. We showed in S18 that the choice of vertical boundary conditions only slightly change the final thermal equilibrium with zero net-flux. Since periodic boundary conditions are much more computationally consuming than modified outflow boundaries, we chose to use the latter. The modified outflow boundaries \citep{Brandenburg1995} assume a zero-gradient extrapolation for all hydrodynamic quantities leaving the box and prevent matter from outside from entering the simulation box. The difference with pure outflow conditions is that they impose the magnetic field to be vertical at the z-boundary. We did not investigate other boundary conditions.

The mass in the shearing box may decrease as a result of the outflow boundaries. To avoid this, we normalized the total mass to the initial mass at each time step by multiplying the density by a corrective factor. For simulations with $\beta$ $\approx$ $10^3$ , outflows become important. In the simulation 439F with $\beta\approx10^3$, the disk would be empty after seven orbits if no normalization were used. When $\beta\approx 10^2$, the disk expands considerably because of magnetic pressure and the location of the photosphere is outside of the box. When losses of matter and energy are important, we normalized the pressure by the same factor as for the density to account for the energy loss of the disk through vertical boundaries. When the pressure was normalized, we observed a change of 40\% in the midplane temperature for $\beta\approx10^2$ compared to only 7\% when $\beta\approx10^3$. We ran two simulations with $\beta$ as low as $\approx10^2$ and a normalized pressure, but tried to avoid this configuration otherwise. 

\subsection{Numerical method}
We solved the MHD equations on a 3D Cartesian grid with the conservative Godunov-type code PLUTO  \citep{mignone2009}. We chose a second-order Runge-Kutta time-integration method. Constrained transport ensures that $\nabla\cdot\textbf{B}=0$ up to machine precision. We used the HLLD solver to solve the Riemann problem. The HLLD solver can be used with a general EOS where $\Gamma$ is not a constant. However, when we computed the maximum and minimum propagation speed waves, we used the Davis estimate \citep{davis1988} and assumed that $\Gamma$ is constant on each side of the propagating waves. The Riemann solver is HLLD except where the pressure difference between two adjacent cells exceeds five times the local pressure, the local ratio of thermal pressure over magnetic pressure is lower than $10^{-3}$ or the Alfv\'enic speed is greater than 30, in which case we used the more diffusive solver HLL. In our simulations each cell spent less than 10\% of its time in HLL. To solve the radiative transfer equations, we followed the same implicit scheme as \cite{Flock2013}, except that we used the biconjugate gradient  solver KSPIBCGS \citep{Yang:2002wg} as implemented in PETSC \citep{petsc}, which we find to provide stabler, faster convergence than BiCGSTAB for our application. In order to avoid very small time steps, we used floors of $10^{-6}$ and $5\times 10^{-2}$ of the initial mid-plane values for the density and temperature, respectively. 

\subsection{Initial conditions\label{sec:init}}

The initial flow was Keplerian with a starting constant vertical magnetic field $B_{z\:0}$ of variable intensity. We chose the intensity of the magnetic fields to be 8 G, 2 G, or 0.8 G. This is consistent with a white dwarf with a radius of $10^9$ cm and a dipolar surface magnetic field of 20 000, 5 000, and 2 000 G, respectively, or a stellar companion of $\approx1\:R_\odot$ with a dipolar surface magnetic field of 60, 15, and 6 G, respectively, at a binary separation of $10^{11}$ cm. The choice of our boundary conditions in $x$ and $y$ ensures that the net flux of magnetic field crossing the equatorial plane is constant throughout our simulation. 

We started with an isothermal layer and assumed hydrostatic equilibrium to fix the initial vertical density profile. We let the MRI develop, then triggered radiative transfer after 32 local orbits (time was normalized by the quantity $1/\Omega_0$ , thus one orbital period is equivalent to $2\pi$ in code units) and let the disk equilibrate and reach a quasi-steady state (if there was one). 

For simulations with $\beta\geq10^5$ , the isothermal temperature $T_{c0}$ was set to be approximately the mid-plane temperature found using the vertical structure code of  \citet{hameury1998} for a given surface density $\Sigma_0$ and a given effective temperature $T_{\rm eff}$ , as in the case of ZNF in S18. This code solves the vertical structure equations assuming an $\alpha$-prescription for the angular momentum transport and associated heating rate. It also uses radiative transfer in the diffusion approximation, and convection described by mixing length theory with a mixing coefficient $\alpha_{\rm ml}=1.5$ (based on models of the Sun). However, when $\beta<10^5$, to avoid inappropriate box sizes that are due to the influence of the magnetic field, we evaluated the expected mid-plane temperature from test simulations. 

\subsection{Runs and diagnostics\label{sec:diag}}

\begin{table*}[!htbp]
\thispagestyle{empty}
\resizebox*{\textwidth}{0.97\textheight}{%
\begin{tabular}{ c | c l c | c | c | c | c | c | c | c | c | c | c | c }
\midrule
 Run & $\Sigma_0$ & $T_{c0}$& $\{T_{mid}\}$ $\pm$ $\sigma_{T_{mid}}$ & $\{T_{eff}\}$ $\pm$ $\sigma_{T_{eff}}$ & $\{\beta\}$ & $\alpha$ $\pm$ $\sigma_{\alpha}$ & $q_\mathrm{surface}$ & $\zeta$ & $f_\mathrm{conv}$ & $\mathrm{\dot{M}_{R\phi}}$ & $\mathrm{\dot{M}_{Z\phi}}$ & $\mathrm{H/R}$ & $t_{avg}$\\
\midrule
\multicolumn{14}{c}{\bfseries{Net Flux Simulations $B_0=8\:G$}} \\
\midrule
\multicolumn{14}{c}{Upper branch} \\
\midrule
439F & 540 &75000 & 76708 $\pm$ 3497 & 11869 $\pm$ 1121 & 120079 & 0.046 $\pm$ 0.014 & 81 & 2.07E-4 & 0.003 & 4.78E17 & 1.38E17 & 3.46E-2 & 700\\
437F & 174 & 50000 & 48009 $\pm$ 1578 & 8218 $\pm$ 424 & 30236 & 0.055 $\pm$ 0.009 & 48 & 2.97E-4 & 0.012 & 1.10E17 & 8.26E16 & 2.74E-2 & 600\\
442F & 113 & 34500 & 38085 $\pm$ 3389 & 7520 $\pm$ 339 & 17177 & 0.078 $\pm$ 0.017 & 52 & 6.43E-4 & 0.10 & 7.56E16 & 8.90E16 & 2.43E-2 & 600\\
456F & 80 & 35000 & 26024 $\pm$ 4367 & 6989 $\pm$ 310 & 9571 & 0.144 $\pm$ 0.040  & 81 & 8.47E-4 & 0.28 & 5.83E16 & 1.39E17 & 2.02E-2 & 600\\
457F & 70 & 30000 & R & R & R & R & R  & R & R & R & R & R & 200\\
\midrule
\multicolumn{14}{c}{Lower branch} \\
\midrule
401F & 200 & 11000 & R & R & R & R &  R & R & R & R & R & R & 600 \\
435F & 191 & 5000 & 13751 $\pm$ 638 & 5656 $\pm$ 302 & 14202 & 0.067 $\pm$ 0.013 & 70 & 1.89E-3 & 0.59 & 2.80E16 & 1.21E17 & 1.46E-2 & 600\\
436F & 150 & 9000 & 9238 $\pm$ 573 & 4544 $\pm$ 274 & 7817 & 0.089 $\pm$ 0.025 & 37 & 3.74E-3 & 0.21 & 1.37E16 & 6.33E16 & 1.20E-2 & 600 \\
465F & 116 &  6000 & 3925 $\pm$ 300 & 3681 $\pm$ 228 & 3839 & 0.156 $\pm$ 0.027 & 34 & 4.28E-3 & 0.22 & 8.06E15 & 5.84E16 & 7.83E-3 & 600\\
434F & 93 & 5000 & 3290 $\pm$ 163 & 3240 $\pm$ 183 & 2563 & 0.186$ \pm$ 0.034 & 31 & 4.37E-3 & 0.26 & 5.52E15 & 5.40E16 & 7.17E-3 & 600\\
477F & 70 & 3000 & 3005 $\pm$ 172 & 3044 $\pm$ 181 & 1591 & 0.275 $\pm$  0.063 & 31 & 1.01E-2 & 0.32 & 5.15E15 & 5.37E16 & 6.81E-3 & 600\\
476F & 50 & 2500 & 2416 $\pm$ 121 & 2653 $\pm$ 150 & 939 & 0.354 $\pm$ 0.060 & 30 & 1.25E-2 & 0.25 & 3.33E15 & 5.22E16 & 6.12E-3 & 600\\
\midrule
\multicolumn{14}{c}{\bfseries{Net Flux Simulations $B_0=2\:G$}} \\
\midrule
\multicolumn{14}{c}{Upper branch} \\
\midrule
439F & 540 & 75000 & 75531 $\pm$ 2228 & 11541 $\pm$ 785 & 1954988 & 0.040 $\pm$ 0.009 & 377 & 1.28E-4 & 0.003 & 4.10E17 & 3.93E16 & 3.44E-2 & 600\\
437F & 174 & 50000 & 44748 $\pm$ 3496 & 7965 $\pm$ 384 & 478899 & 0.047 $\pm$ 0.015 & 256 & 1.21E-4 & 0.08 & 8.60E16 & 1.81E16 & 2.64E-2 & 600\\
442F & 113 & 34500 & 34349 $\pm$ 5792 & 7386 $\pm$ 327 & 263340 & 0.078 $\pm$ 0.027 & 264 & 2.66E-4 & 0.20 & 6.67E16 & 2.75E16 & 2.32E-2 & 600\\
452F & 100 & 30000 & 28530 $\pm$ 4463 & 7064 $\pm$ 293 & 207469 & 0.097 $\pm$ 0.025 & 322 & 3.06E-4 & 0.37 & 5.62E16 & 3.37E16 & 2.11E-2 & 600\\
454F & 90 & 28000 & R & R & R & R & R & R & R  & R & R & R & 150\\
\midrule
\multicolumn{14}{c}{Middle branch} \\
\midrule
415F & 360 & 14000 & 14194 $\pm$ 298 & 5913 $\pm$ 402 & 552927 & 0.046 $\pm$ 0.018 & X & X & 0.81 & 4.32E16 & X & 1.49E-2 & 600\\
413F & 340 & 10000 & 12259 $\pm$ 127 & 5230 $\pm$ 243 & 380814 & 0.041 $\pm$ 0.011 & X & X & 0.61 & 2.50E16 & X & 1.38E-2 & 600\\
411F & 320 & 4500 & 9840 $\pm$ 113 & 4140 $\pm$ 419 & 291836 & 0.037 $\pm$ 0.012 & X & X & 0.74 & 1.39E16 & X & 1.24E-2 & 600\\
\midrule
\multicolumn{14}{c}{Lower branch} \\
\midrule
409F & 300 & 4000 & 4973 $\pm$ 842 & 3497 $\pm$ 156 & 187517 & 0.028 $\pm$ 0.010 & 74 & 8.84E-4 & 0.31 & 4.69E15 & 7.72E15 & 8.78E-3 & 1500\\
407F & 280 & 4000 & 3450 $\pm$ 71 & 2996 $\pm$ 137 & 128410 & 0.028 $\pm$ 0.009 & 63 & 2.75E-4 & 0.16 & 2.46E15 & 6.61E15 & 7.31E-3 & 600\\
404F & 230 & 3500 & 3465 $\pm$ 92 & 3053 $\pm$ 147 & 106544 & 0.039 $\pm$ 0.014 & 85 & 6.70E-4 & 0.15 & 2.93E15 & 8.85E15 & 7.32E-3 & 600 \\
403F & 220 & 3500 & 3230 $\pm$ 70 & 2925 $\pm$ 135 & 94363 & 0.038 $\pm$ 0.011 & 67 & 4.44E-4 & 0.09 & 2.31E15 & 6.99E15 & 7.07E-3 & 600 \\
401F & 200 & 4000 & 2728 $\pm$ 142 & 2672 $\pm$ 137 & 74292 & 0.035 $\pm$ 0.009 & 49 & 1.94E-4 & 0.03 & 1.43E15 & 5.10E15 & 6.51E-3 & 600 \\
465F & 116 & 3000 & 2176 $\pm$ 98 & 2371 $\pm$ 130 & 37650 & 0.051 $\pm$ 0.011 & 50 & 3.91E-4 & 0.03 & 9.97E14 & 5.20E15 & 5.80E-3 & 600\\
434F & 93 & 2500 & 2021 $\pm$ 107 & 2243 $\pm$ 144 & 28981 & 0.055 $\pm$ 0.014 & 41 & 4.96E-4 & 0.03 & 7.69E14 & 4.34E15 & 5.60E-3 & 600\\
477F & 70 & 2000 & 1885 $\pm$ 44 & 2097 $\pm$ 125 & 21167 & 0.065 $\pm$ 0.012 & 45 & 9.96E-4 & 0.03 & 6.47E14 & 4.78E15 & 5.42E-3 & 600\\
\midrule
\multicolumn{14}{c}{\textbf{Net Flux Simulations $B_0=0,8\:G$}} \\
\midrule
\multicolumn{14}{c}{Upper branch} \\
\midrule
439F & 540 & 75000 & 73173 $\pm$ 3826 & 10731 $\pm$ 1269 & 12025687 & 0.032 $\pm$ 0.013 & 2116 & 8.09E-5 & 0.001 & 3.17E17 & 3.53E16 & 3.38E-2 & 600\\
437F & 174 & 50000 & 47560 $\pm$ 3365 & 8367 $\pm$ 528 & 3068497 & 0.054 $\pm$ 0.017 & 1121 & 1.52E-4 & 0.019 & 1.06E17 & 1.87E16 & 2.73E-2 & 900\\
442F & 113 & 35000 & 31312 $\pm$ 6879 & 7273 $\pm$ 316 & 1561523 & 0.082 $\pm$ 0.025 & 1099 & 2.51E-4 & 0.27 & 6.22E16 & 1.83E16 & 2.21E-2 & 800\\
452F & 100 & 30000 & R & R & R & R & R & R & R & R & R & R & R \\
\midrule
\multicolumn{14}{c}{Middle branch} \\
\midrule
415F & 360 & 5000 & 10123 $\pm$ 245 & 4259 $\pm$ 187 & 2117377 & 0.031 $\pm$ 0.009 & X & X & 0.83 & 1.41E16 & X & 1.26E-2 & 600\\
\midrule
\multicolumn{14}{c}{Lower branch} \\
\midrule
411F &  320 & 4000 & 3480 $\pm$ 101 & 3002 $\pm$ 112 & 1140102 & 0.024 $\pm$ 0.008 & 148 & 1.89E-4 & 0.15 & 2.71E15 & 2.48E15 & 7.35E-3 & 1200\\
409F &  300 & 3500 & 3623 $\pm$ 54 & 3093 $\pm$ 135 & 911231 & 0.029 $\pm$ 0.009 & 202 & 4.41E-4 & 0.18 & 3.08E15 & 3.38E15 & 7.48E-3 & 600\\
407F &  280 & 3000 & 3508 $\pm$ 92 & 3026 $\pm$ 113 & 814922 & 0.032 $\pm$ 0.011 & 212 & 6.30E-4 &  0.19 & 2.93E15 & 3.54E15 & 7.37E-3 & 1000\\
405F &  250 & 2500 & 3209 $\pm$ 131 & 2864 $\pm$ 155 & 665472 & 0.038 $\pm$ 0.017 & 338 & 8.89E-4 & 0.14 & 2.60E15 & 5.64E15 & 7.07E-3 & 600 \\
404F &  230 & 2000 & 2810 $\pm$ 118 & 2626 $\pm$ 131 & 540606 & 0.035 $\pm$ 0.010 & 152 & 7.82E-4 & 0.1 & 1.72E15 & 2.55E15 & 6.62E-3 & 600 \\
403F &  220 & 3500 & 2788 $\pm$ 512 & 2689 $\pm$ 312 & 531776 & 0.035 $\pm$ 0.014 & 165 & 2.51E-4 &  0.05 & 1.73E15 & 2.75E15 & 6.57E-3 & 1500 \\
401F &  200 & 3500 & 2716 $\pm$ 92 & 2667 $\pm$ 129 & 462249 & 0.037 $\pm$ 0.010 & 155 & 2.35E-4 & 0.04 & 1.55E15 & 2.60E15 & 6.48E-3 & 600 \\
465F &  116 & 3000 & 1954 $\pm$ 92 & 2170 $\pm$ 135 & 224681 & 0.035 $\pm$ 0.009 & 93 & 1.57E-4 &  0.01 & 6.01E14 & 1.56E15 & 5.50E-3 & 600\\
434F &  93 & 2000 & 1878 $\pm$ 45 & 2037 $\pm$ 146 & 175993 & 0.039 $\pm$ 0.011 & 99 & 3.59E-4 & 0.02 & 5.16E14 & 1.65E15 & 5.41E-3 & 600\\
477F &  70 & 2000 & 1806 $\pm$ 18 & 1877 $\pm$ 106 & 130298 & 0.041 $\pm$ 0.007 & 98 & 3.74E-4 & 0.02 & 3.81E14 & 1.63E15 & 5.31E-3 & 600\\
\midrule
\multicolumn{14}{c}{\textbf{Net Flux Simulations} $\beta\approx10^4$} \\
\midrule
\multicolumn{14}{c}{Upper branch} \\
\midrule
439F & 540 & 79000 & 82633 $\pm$ 3629 & 13426 $\pm$ 1143 & 12166 & 0.077 $\pm$ 0.020 & 33 & X & 0.01 & 8.53E17 & 5.73E17 &3.59E-2 & 950\\
442F & 113 & 34500 & 39219 $\pm$ 2659 & 7767 $\pm$ 444 & 12732 & 0.086 $\pm$ 0.018 & 46 & X & 0.07 & 8.83E16 & 1.10E17 & 2.48E-2 & 600\\
452F & 100 & 34500 & 37777 $\pm$ 2428 & 7523 $\pm$ 388 & 12122 & 0.093 $\pm$ 0.021 & 55 & X & 0.11 & 7.72E16 & 1.15E17 & 2.4E-2 & 600\\
454F & 90 & 34500 & 30721 $\pm$ 6055 & 7200 $\pm$ 369 & 10793 & 0.110 $\pm$ 0.034 & 55 & X & 0.17 & 6.48E16 & 1.05E17 & 2.18E-2 & 600\\ 
456F & 80 & 34500 & 29261 $\pm$ 4841 & 7221 $\pm$ 346 & 10470 & 0.128 $\pm$ 0.033 & 71 & X & 0.20 & 6.18E16 & 1.21E17 & 2.14E-2 & 550\\ 
\midrule
\multicolumn{14}{c}{Lower branch} \\
\midrule
434F & 93 & 1976 & 2392 $\pm$ 70 & 2608 $\pm$ 116 & 7168 & 0.099 $\pm$ 0.015 & 34 & X & 0.05 & 1.68E15 & 1.58E16 & 6.11E-3 & 600\\
434F\_$\beta$12000 & 93 & 1976 & 2109 $\pm$ 38 & 2376 $\pm$ 101 & 13450 & 0.067 $\pm$ 0.009 & 46 &  X & 0.03 & 9.94E14 & 1.07E16 & 5.74E-3 & 600\\
477F & 70 & 1976 & 3065 $\pm$ 193 & 2967 $\pm$ 148 & 9063 & 0.104 $\pm$ 0.019 & 73 &  X & 0.05 & 1.96E15 & 2.56E16 & 6.92E-3 & 600\\
\midrule
\multicolumn{14}{c}{\textbf{Net Flux Simulations} $\beta\approx10^3$} \\
\midrule
\multicolumn{14}{c}{Upper branch} \\
\midrule
439F & 540 & 79000 & 93180 $\pm$ 13750 & 19550 & 2791 & 0.424 $\pm$ 0.126 & 36 & X & 0.06 & 5.35E18 & 6.56E18 & 3.82E-2 & 300\\
442F & 113 & 50000 & 49332 $\pm$ 6163 & 12031 $\pm$ 1274 & 637 & 0.444 $\pm$ 0.082 & 44 & X & 0.04 & 6.02E17 & 1.23E18 & 2.78E-2 & 520\\
452F & 100 & 45000 & 45917 $\pm$ 4835 & 11569 $\pm$ 1167 & 490 & 0.469 $\pm$ 0.101 & 45 & X & 0.07 & 5.27E17 & 1.09E18 & 2.68E-2 & 380\\
\midrule
\multicolumn{14}{c}{Lower branch} \\
\midrule
434F & 93 & 4500 & 11915 $\pm$ 4413 & 5993 $\pm$ 529 & 1051 & 0.315 $\pm$ 0.097 & 61 & X & 0.88 & 4.84E16 & 2.98E17 & 1.36E-2 & 500\\
\midrule
\multicolumn{14}{c}{\textbf{Net Flux Simulations} $\beta\approx10^2$} \\
\midrule
\multicolumn{14}{c}{Upper branch} \\
\midrule
442F & 113 & 34500 & 31015 $\pm$ 9461 & 10821 $\pm$ 2124 & 101.9 & 0.850 $\pm$ 0.191 & 32 & X & X & 8.03E17 & 3.78E18 & 2.20E-2 & 200\\
\midrule
\multicolumn{14}{c}{Lower branch} \\
\midrule
434F & 93 & 2000 & 5926 $\pm$ 2007 & 4609 $\pm$ 662 & 153.1 & 0.774 $\pm$ 0.280 & 32 & X & X & 5.32E16 & 7.41E17 & 9.62E-3 & 200\\
\bottomrule
\end{tabular}}
\caption{Initial parameters and results for our ideal MHD simulations. $\Sigma_0$  is the initial surface density in g\,cm$^{-2}$, $T_{c0}$ the initial midplane  temperature in K, and $H/R_0$ where $H$ is the corresponding scale height. Brackets $\{\}$ are for quantities averaged over $t_\mathrm{avg}$ (given in local orbits) with $\sigma$ the associated standard deviation. R signals simulations with runaway heating or cooling. 
}
\label{tab}
\end{table*}

Table \ref{tab} lists the runs that we performed. We partly adopted the notation of \cite{Hirose2014}  to label the runs. $\Sigma_0$ is the initial surface density and $H\equiv c_s(T_{c0})/\Omega$ is the pressure scale-height (with $c_s$ the sound speed). The horizontal extent of the box was $\pm$6$H$ for the hot branch and $\pm$3$H$ for the cold and middle branch. $L_x,\:L_y$ , and $L_z$ follow the ratio 1:4:8 on the hot branch and 1:4:4 on the cold branch. The resolution was $32\times128\times256,$ except for test simulations. We chose to use smaller vertical boxes on the cold branch to avoid high Alfv\'en velocities near the boundaries, resulting in an enhanced numerical diffusion and thus numerical heating.

In the following, we use several averaged quantities denoted in the following way:
\begin{gather*}
\braket{X}_{\rho}\equiv\frac{1}{\Sigma L_xL_y}\int_{-L_z/2}^{L_z/2}\int_{-L_y/2}^{L_y/2}\int_{-L_x/2}^{L_x/2}\rho X\:dxdydz\\
\braket{X}_{x,y,z}\equiv\frac{1}{L_xL_yL_z}\int_{-L_z/2}^{L_z/2}\int_{-L_y/2}^{L_y/2}\int_{-L_x/2}^{L_x/2}X\:dxdydz\\
\braket{X}\equiv\frac{1}{L_xL_y}\int_{-L_y/2}^{L_y/2}\int_{-L_x/2}^{L_x/2}X\:dxdy,
\end{gather*}
where 
\begin{equation*}
\Sigma=\int_{-L_z/2}^{L_z/2}\braket{\rho}_{x,y} dz.
\end{equation*}

We use $\text{square}$ brackets to denote spatial averages and $\text{curly}$ brackets to indicate an average  over a time $t_\mathrm{avg}$ performed when the simulation reached a quasi-steady state. The averaging timescale $t_\mathrm{avg}$ is indicated for each run in the last column of Table \ref{tab}. We typically averaged over 100 orbits.  $\{\Sigma\}$, $\{T_\mathrm{mid}\},$ and $\{T_\mathrm{eff}\}$ are thus the time-averaged values of the surface density, midplane temperature, and effective temperature, respectively. $\sigma_{T_\mathrm{mid}}$, $\sigma_{T_\mathrm{eff}}$ , and $\sigma_{\alpha}$ are the standard deviations of the fluctuations with time of these quantities.

We computed $T_\mathrm{eff}$ as
\begin{equation*}
T_\mathrm{eff}^4=\frac{1}{2\sigma_B}(F_\mathrm{rad\:z}^+-F_\mathrm{rad\:z}^-),
\end{equation*}
where $F_\mathrm{rad\:z}^{+/-}$ is the radiative flux in the vertical direction on the upper or lower boundary of our simulation box.
We also defined $\braket{Q^-}_{x,y}$ the local cooling rate as
\begin{equation}
\braket{Q^-}_{x,y}\equiv \frac{d}{dz}\braket{F_\mathrm{rad\:z}}_{x,y}+\frac{d}{dz}\braket{\epsilon v_z}_{x,y}
\label{eq:fluxes}
.\end{equation}
$v_z$ is the vertical velocity, and the quantity $\epsilon v_z$ represents the advective flux of internal energy $F_\mathrm{adv}$. 

We defined $\alpha$, the ratio of stress to pressure, as
\begin{equation*}
\alpha \equiv \frac{\{\braket{W_{xy}}_{x,y,z}\}}{\{\braket{P}_{x,y,z}\}},
\end{equation*}
and $\beta$, the ratio of the time-averaged midplane pressure to the magnetic pressure due to the mean vertical field as follows:
\begin{equation} \label{eq:beta}
\beta = \frac{8\pi\{\braket{P_{\mathrm{thermal\:mid}}}\}}{\braket{B_\mathrm{z}}^2}
.\end{equation}

The instantaneous stress-to-pressure ratio is $\tilde{\alpha}=\braket{W_{xy}}_{x,y,z}/\braket{P}_{x,y,z}$.
The turbulent stress $W_{xy}$ is $\rho (u_xu_y-v_{\mathrm{A}x}v_{\mathrm{A}y}),$ where $\pmb{u}=\pmb{v}+\frac{3}{2}\Omega x\pmb{\hat{y,}}$ and $v_\mathrm{A}$ is the Alfv\'en velocity computed from the mean field.

\section{Ideal MHD runs}
First, we gather all of our simulations and discuss the influence of a net vertical magnetic field on the local properties of MRI in \S\ref{sec:local} and the properties of MRI-driven outflows in \S\ref{sec:outflows}. In \S\ref{sec:Scurve}, we discuss the impact of a large-scale poloidal magnetic field with three different magnitudes on the S curves. In \S\ref{sec:accretion}, we investigate the relative importance of viscous and wind transport of angular momentum on the ensuing accretion rate. Finally, in \S\ref{sec:DIM_wind}, we discuss the impact of the wind-driven angular momentum transport on observables.

\subsection{Local turbulent transport\label{sec:local}}
\begin{figure}[h!]
\includegraphics[width=100mm,height=60mm]{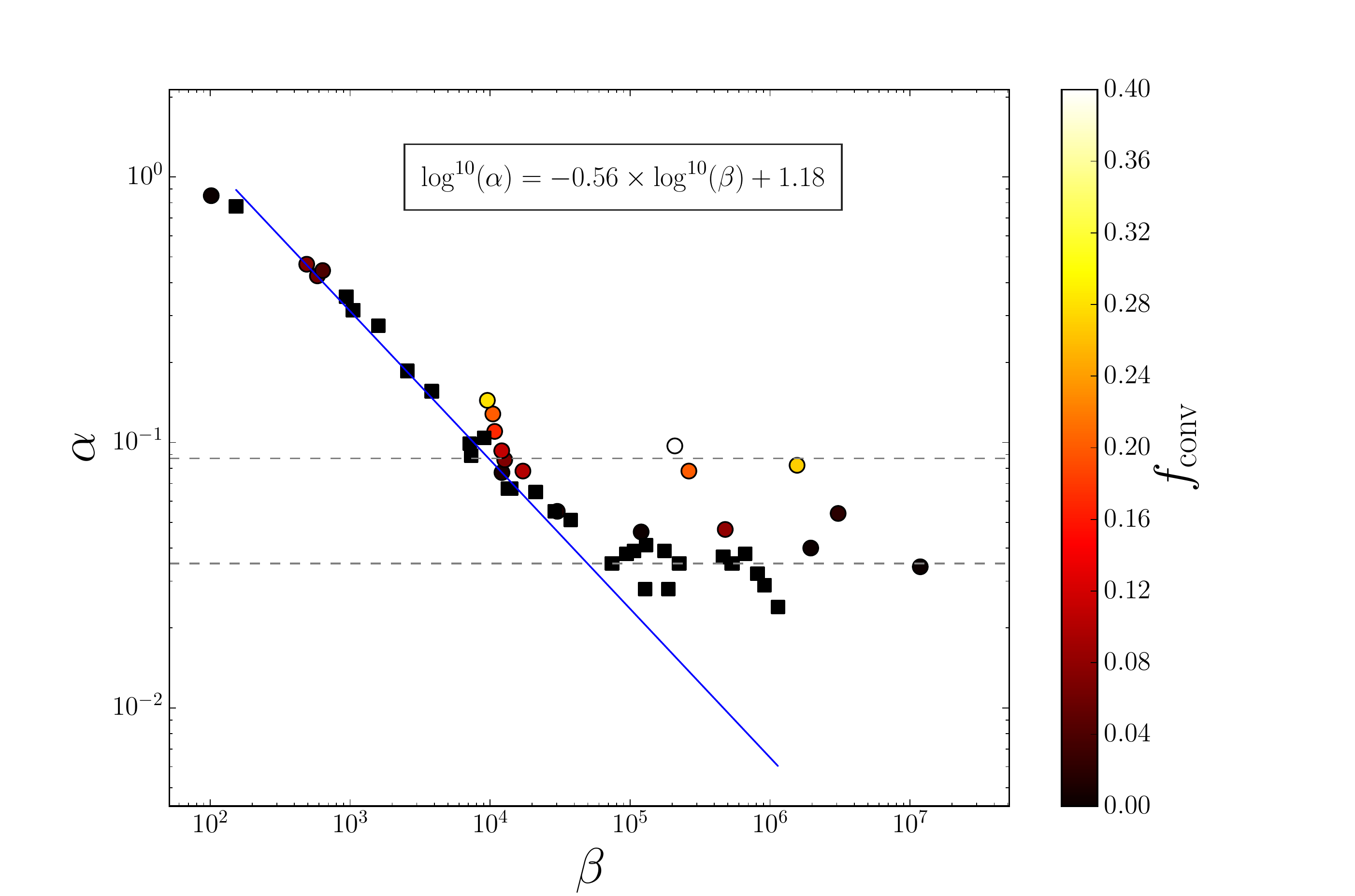}
\caption{Square markers denote cold branch simulations, and circle markers denote hot branch simulations. The color indicates the value of $\mathrm{f_{conv}}$. Horizontal dotted gray lines show the value of $\alpha$ in ZNF simulations from S18 in a highly convective simulation of the hot branch (upper line) and a typical cold branch case (bottom line). The linear fit is made using data from the cold branch simulations with $\beta\lesssim2\times10^4$.}
\label{alpha_beta}
\end{figure}

Local MRI turbulent transport is well known to depend on the magnetization parameter $\beta$ \citep{Hawley1995}. We plot in Figure \ref{alpha_beta} $\alpha$ as a function of $\beta$ for all of our simulations. Squares are used for cold branch simulations and circles for the hot branch (see \S\ref{sec:Scurve} for details on the cold and hot branch). The color of the circles gives $f_\mathrm{conv}$ , which is defined as

\small
\begin{equation*}
f_\mathrm{conv} \equiv \frac{1}{2}(\mathrm{max}(0,\mathrm{max}(\big\{\frac{F_\mathrm{adv}(z>0)}{F_\mathrm{rad\:z}^+}\big\}))+  \mathrm{max}(0,\mathrm{max}(\big\{\frac{F_\mathrm{adv}(z<0)}{F_\mathrm{rad\:z}^-}\big\}))),
\end{equation*}
\normalsize
where $F_\mathrm{adv}=\braket{ev_z}$. $f_\mathrm{conv}$ is a dimensionless measure of the importance of convective transport compared to radiative transport.

The gray dashed lines in Figure \ref{alpha_beta} correspond to the value of $\alpha$ for the convective run 452O, which shows an enhanced $\alpha$ (top), and for the purely radiative run 434O (bottom) in S18. $\alpha$ ranges between 0.275 and 0.026. We find that $\alpha$ scales as $\beta^{-0.56}$ (with $\beta$ defined in Eq.(\ref{eq:beta})) when we only include the simulations from the cold branch. This is consistent with previous works (\citealt{Hawley1995}; \citealt{salvessen2016}). 

As in ZNF, convection plays a role in determining the value of $\alpha$ (see \cite{Hirose2014} and S18). For $\beta>10^5$, simulations from both the hot and cold branch depart from the $\beta^{-0.56}$ relation and tend toward the ZNF case. Moreover, Figure \ref{alpha_beta} shows for $\beta\approx10^4$  that convective simulations from the hot branch leave the relation, whereas cold branch simulations do not. We note an increasing deviation from the trend with $f_\mathrm{conv}$ for these convective simulations. However, in simulations with a high level of magnetization, the disk is thicker because of the magnetic pressure in the midplane and thus pushes the zone where convection could set in out of the box. We doubled the vertical size and resolution of the disk in 456F\_$\mathrm{bigbox}\_\beta3,$ but only observed one convection episode. The effect on $\alpha$ is not clear from this single event. All we can say is that the vertical equilibrium is dominated by magnetic pressure, hence it would seem reasonable that convection plays a lesser role than that for higher $\beta$. We add the caveat that, in this particular simulation, pressure is normalized. This may have an impact on the onset of convection.

\begin{figure}[h!]
\includegraphics[width=105mm,height=75mm]{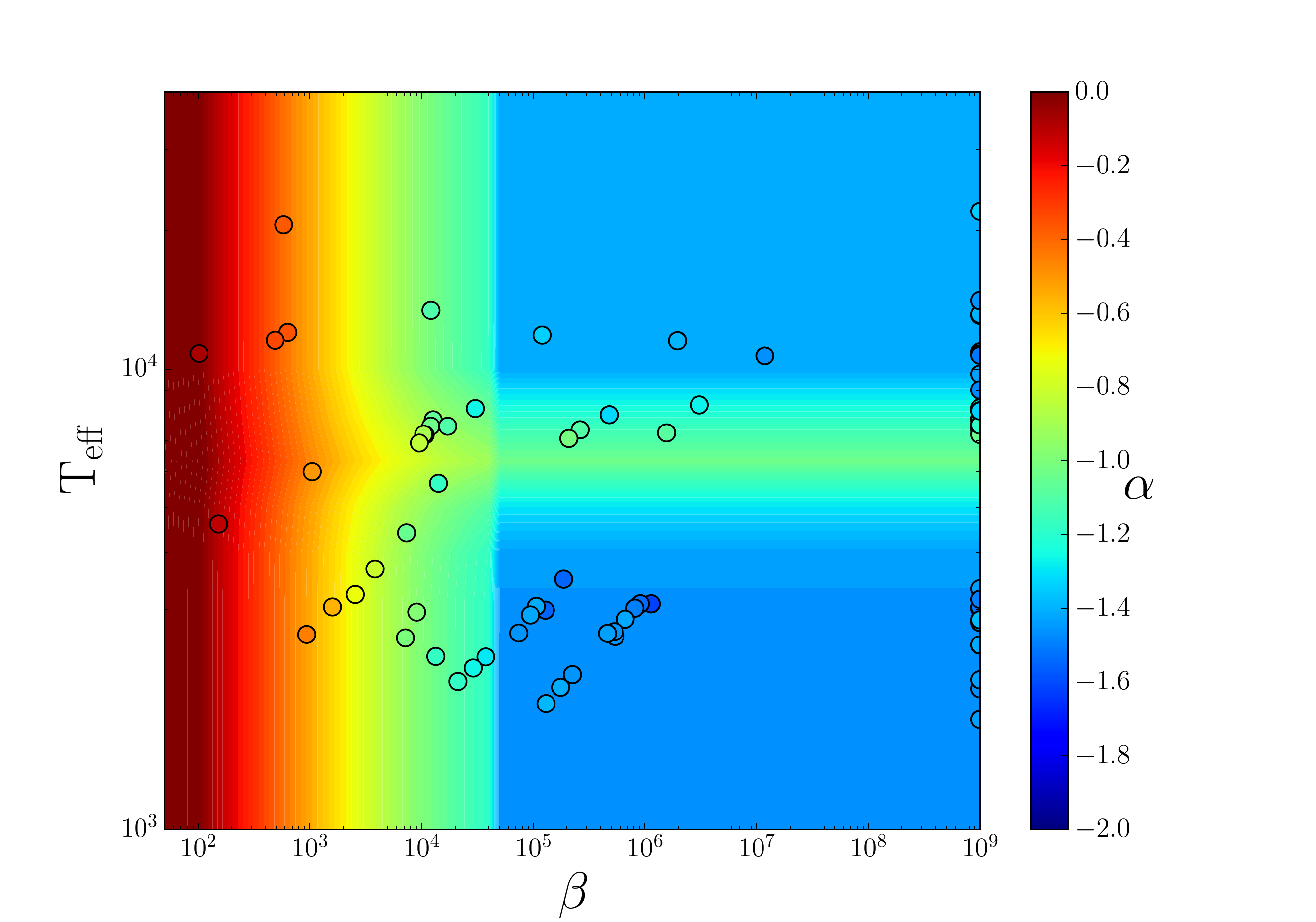}
\caption{Value of $\log(\alpha)$ in our simulations is shown by color. Simulations at $\beta=10^9$ are ZNF simulations taken from S18 with additional simulations from this article. The background color map shows the result of the fit of $\alpha$ as a function of $\mathrm{T_{eff}}$ and $\beta$ from Eq. (\ref{eq:fit}).}
\label{fit_alpha}
\end{figure}

S18 showed that $\alpha$ is not correlated with the convective fraction. This result was also confirmed by \cite{coleman2018}, who observed that $\alpha$ appears to be more correlated with the vertical Mach number. However, they added that the speed of the convective eddies is only 10\% of the turbulent speed. This indicates that convection is weaker than turbulence and does not significantly drive additional motion that could feed the dynamo. All these characteristics are also true for the net flux. In the absence of any deep insight on how convective MRI leads to an enhancement of $\alpha$, we plot in Figure \ref{fit_alpha} a 2D map of $\alpha$ in the $\mathrm{T_{eff}}$, $\beta$ space. We fit our data with Eq. (\ref{eq:fit}), which is similar to \cite{coleman2016} with an additional term for the dependence in $\beta$ based on Figure \ref{alpha_beta},

\small
\begin{equation} \label{eq:fit}
\alpha= \begin{cases} A\exp(-\frac{(T_\mathrm{eff}-T_0)^2}{2\sigma^2})+B\tanh(\frac{T_\mathrm{eff}-T_0}{\sigma})+C+D\beta^{-E}, & \mbox{if } \beta<5\times10^4 \\ A\exp(-\frac{(T_\mathrm{eff}-T_0)^2}{2\sigma^2})+B\tanh(\frac{T_\mathrm{eff}-T_0}{\sigma})+C, & \mbox{if } \beta>5\times10^4 \end{cases} 
.\end{equation}
\normalsize
We chose to constrain A, $T_0$, $\sigma$, B, and C using only simulations with $\beta>5\times10^4$. Then, we fit the remaining parameters in the simulations with $\beta<10^4$. We plot the fit on the background of Figure \ref{fit_alpha}. The best-fit parameters are $A=5.35\times10^{-2}$, $T_0=6866\:K$, $\sigma= 853\:K$, $B=1.65\times10^{-3}$, $C=3.65\times10^{-2}$, $D=1.37,$ and $E=0.58$. This prescription will be useful in a DIM model or any global model that does not resolve the turbulent transport.

\subsection{Outflows and angular momentum transport\label{sec:outflows}}

Net magnetic flux significantly affects the dynamic of the disk as it allows the formation of magneto-centrifugally driven outflows from MRI (\citealt{fromang2013}; \citealt{lesur2013}). All of our simulations with a $\beta\lesssim10^4$ clearly exhibit these magneto-centrifugally driven outflows. By computing the Bernoulli invariant along streamlines, we observe that thermal effects also play a significant role in launching the outflow. Our simulations reproduce the main characteristic that have been extensively described in the literature (\citealt{blandford1982}, \citealt{ferreira1995}, \citealt{fromang2013}; \citealt{lesur2013}).

Such outflows extract matter, angular momentum, and energy away from the disk. Since the classical DIM only includes transport of angular momentum through anormalous diffusive transport, we need to quantify the transport that is due to the outflows to include them in a more sophisticated DIM. Following \cite{balbus1999}, we derived the conservation of angular momentum of an axisymmetric disk in cylindrical coordinates \textit{\textup{including}} the surface terms. After some algebra, we can write the surface density equation of evolution in cylindrical coordinates in the following form:

\tiny
\begin{equation}\label{eq:DIM}
\partial_t\Sigma-\frac{2}{R_0}\partial_{R_0}\left(R_0^{1/2}\partial_{R_0}\left(\frac{\Sigma}{\Omega_0} R_0^{1/2}\alpha \braket{c_s^2}_{\rho}\right)+\frac{2q}{\beta}\braket{P_\mathrm{thermal\:mid}}\frac{R_0}{\Omega_0}\right)+\zeta\braket{\rho_\mathrm{mid}c_\mathrm{s\:mid}}=0
.\end{equation}
\normalsize
Equation (\ref{eq:DIM}) shows three ways for the surface density to change: through radial accretion due to a transport of angular momentum either through turbulence or outflows, and loss of matter from the vertical boundaries. We have studied the radial turbulent transport of angular momentum defined by the $\alpha$ parameter in \S\ref{sec:local}. Outflows lead to the definition of two additional dimensionless parameters: $q,$ which quantifies the strength of the wind torque, and $\zeta,  $ which is the ratio of the mass loss from the vertical boundaries to the sound speed times the density in the midplane. They are defined as 
\begin{equation}
q=\frac{[\braket{\rho u_{\phi}u_z} -\braket{B_\mathrm{\phi}B_z}/4\pi]_\mathrm{surface-}^\mathrm{surface+}}{2\braket{P_\mathrm{thermal\:mid}}}\beta\
\end{equation}
and
\begin{equation}
\zeta=\frac{\left[\braket{\rho u_z}\right]_{-L_z/2}^{L_z/2}}{2\braket{\rho_\mathrm{mid}c_\mathrm{s\:mid}}}.
\end{equation}
Shearing box simulations are very useful for studying the local properties of MRI turbulence. However, they suffer limitations concerning the global geometry of the outflows. The shearing-box symmetries do not ensure a physical configuration of the magnetic field. For example, they can allow a surface torque of the same sign on both sides of the disk; this configuration does not extract angular momentum. Additionally, it is a well-known result that the sign of $\braket{B_y}$ flips regularly in a shearing box \citep{Brandenburg1995}. To cope with these limitations, we defined a function $S$ that gives the sign of $\braket{B_y}$ in each hemisphere. We multiplied the wind torque by $S$ in our time average. This allowed us to compute wind torques that do not depend on the geometry picked up by the shearing box at a given time.

A surface for the disk needs to be defined so that $q$ can be estimated. We took the height where the maximum value of the wind torque is reached as the definition of the disk surface because above this surface, magnetic energy is transferred to kinetic energy to accelerate the outflow. We add the caveat that our vertically modified outflow boundary conditions impose a purely vertical magnetic field that constrains the geometry of the magnetic field lines and therefore imposes $q=0$ at the boundaries. Numerically, we defined $q$ as

\small
\begin{equation}
q_\mathrm{surface}=\frac{\mathrm{max}\big(S(\braket{\rho v_{\phi}v_z}+\braket{B_\mathrm{\phi}B_z})(z^+)+S(\braket{\rho v_{\phi}v_z}+\braket{B_\mathrm{\phi}B_z})(z^-)\big)}{2\braket{P_\mathrm{thermal\:mid}}}\beta
.\end{equation}
\normalsize

\begin{figure}[h!]
\includegraphics[width=105mm,height=70mm]{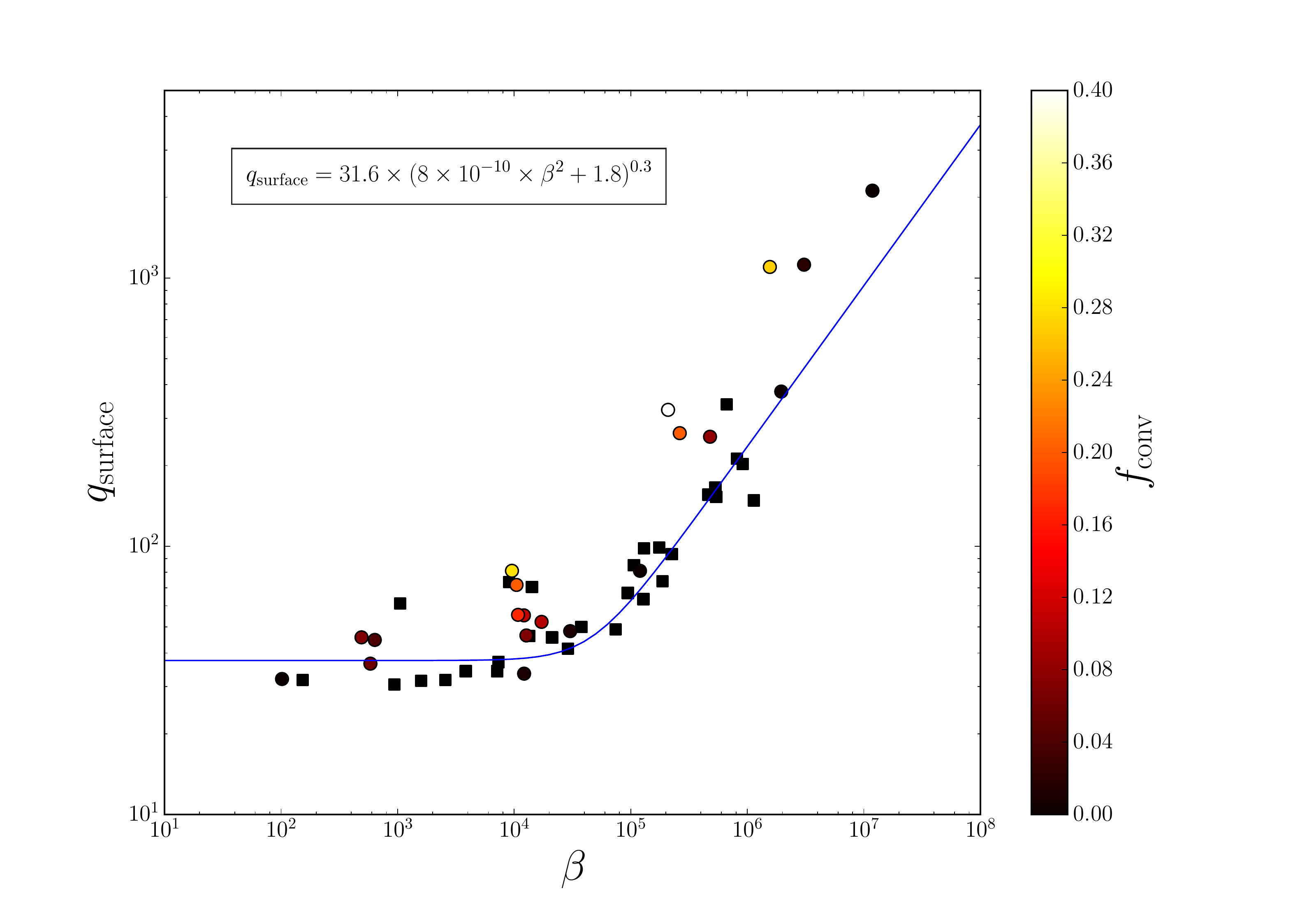}
\caption{Squares denote cold branch simulations, and circles denote hot branch simulations. The color indicates the value of $\mathrm{f_{conv}}$.}
\label{q_over_beta_vs_beta}
\end{figure}

We plot $q_\mathrm{surface}$ in Figure \ref{q_over_beta_vs_beta} as a function of $\beta$ for all simulations. We find that $q_\mathrm{surface}$ is roughly constant for $\beta\lesssim10^4$ with an approximate value of $q_\mathrm{surface}\approx40$. For low $\beta,$ the wind torque is dominated by the magnetic part. This means that, remarkably, across almost three orders of magnitude in $\beta$, the MRI gives a quasi-constant ratio $B_{\phi}/B_z$. This value agrees with local simulations from \cite{fromang2013} and \cite{simon2013a} as well as with the global simulation of \cite{zhu2018}, where $q\approx10,$  according to their Figures 6 and 7. \cite{bai2013a} found a lower value of $q$, but they measured the wind torque at the external boundaries; our results again agree when we adopt the same definition. When $\beta$ exceeds $10^5$, we tend toward the ZNF case where the magnetic field becomes mainly azimuthal, and we expect $q\propto\beta^{0.5}$. We find that a dependence in $\beta^{0.6}$ fits our data well. We propose in Figure \ref{q_over_beta_vs_beta} a function that allows a smooth transition between the two asymptotic regimes. 

The amplitude of the wind torque normalized by thermal pressure is given by $2q_\mathrm{surface}/\beta$. It varies as $\beta^{-1}$ for $\beta\lesssim10^4$ and as $\beta^{-0.4}$ for lower magnetization; angular momentum is more efficiently removed by the wind at low $\beta,$ as we expected (see \S\ref{sec:accretion}).

Convective simulations from the hot branch have a slightly higher $q_\mathrm{surface}$ than cold branch simulations for $\beta>10^5$. It is unclear whether convection affects the $\phi z$ component of the stress tensor for higher magnetization. We note that the magnetic component $\braket{B_{\phi}B_z}$ accounts for $\approx70\%$ of the wind torque for $\beta\lesssim10^5$. For $\beta\gtrsim10^5$, the Reynolds component can account for up to $\approx60\%$ of the total wind torque.

Since the vertical extent of our box is only a few scale heights, part of the matter leaving the box, which in our simulations is definitely lost, is susceptible to falling back onto the disk. This picture is consistent with decreasing mass outflow with increasing box size, as observed in \cite{fromang2013}; this is also observed here. The value of $\zeta$ can therefore only be an estimate in a shearing-box simulation (\citealt{suzuki2009};  \citealt{fromang2013}). We computed the critical points for the simulation 434F\_8G and found that the fast Alfv\'en point is reached at the vertical boundary exactly as in \cite{fromang2013}, suggesting a numerical artifact of the shearing box. Taking this into account, we conclude that the $\zeta$ we measured are maximum values for the mass loss rate through the vertical boundaries.

\begin{figure}[h!]
\includegraphics[width=100mm,height=65mm]{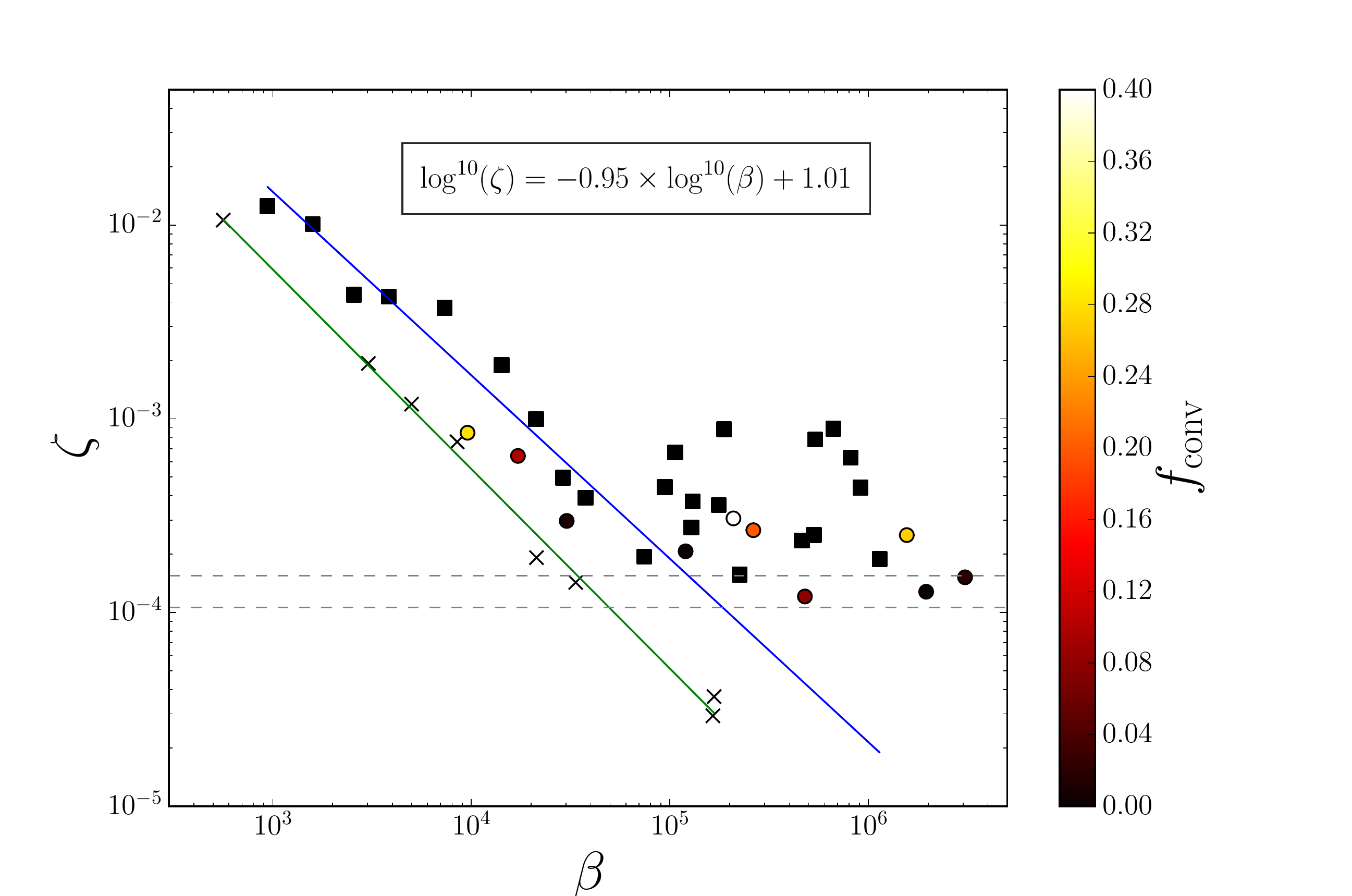}
\caption{Squares denote cold branch simulations with a vertical extent $\mathrm{L_z}=6H,$ and circles denote hot branch simulations with a vertical extent $\mathrm{L_z}=12H$. Crosses are used for cold branch simulations with the same box size as on the hot branch (results from those simulations are not reported in Table \ref{tab}). The color indicates the value of $\mathrm{f_{conv}}$. Horizontal dotted gray lines show the value of $\zeta$ in ZNF simulations from S18 in a highly convective simulation of the hot branch (upper line) and a typical cold branch case (bottom line). The blue and green solid lines are linear fits to the points from the cold branch simulations with $\beta\lesssim2\times10^4$. This figure only reports results from simulations with $\mathrm{B_{z0}=}$0.8, 2, and 8 G.}
\label{zeta_beta}
\end{figure}

We show in Figure \ref{zeta_beta} the dependence of $\zeta$ on $\beta$. Crosses show simulations on a cold branch with an extent of 12H in the z-direction, and squares show simulations with an extent of 6H in the z-direction, as for Figure \ref{alpha_beta} and \ref{q_over_beta_vs_beta}. The dependence of $\zeta$ on the vertical size of the box is clear. Smaller boxes exhibit higher $\zeta$. However, $\zeta$ scales approximately as $\beta^{-1}$ for the two box sizes in the range $\beta<10^5$. Like $\alpha$ and $q_\mathrm{surface}$, $\zeta$ reaches the asymptotic value given by the zero net-flux case as we increase $\beta$. The results are more dispersed than for $\alpha$ since by definition, they depend on the boundary conditions and the relative box size compared to the disk. All these results are consistent with \cite{suzuki2009}, \cite{bai2013a}, and \cite{fromang2013}. Moreover, it seems that convection enhances $\zeta$ for $\beta>10^5$ , whereas it does not for lower $\beta$. This is illustrated by the higher $\zeta$ of the convective hot branch simulations. 

\subsection{Thermal equilibrium curves with a constant value of $B_{z_{0}}$\label{sec:Scurve}}
\begin{figure*}[h!]
\includegraphics[width=\textwidth]{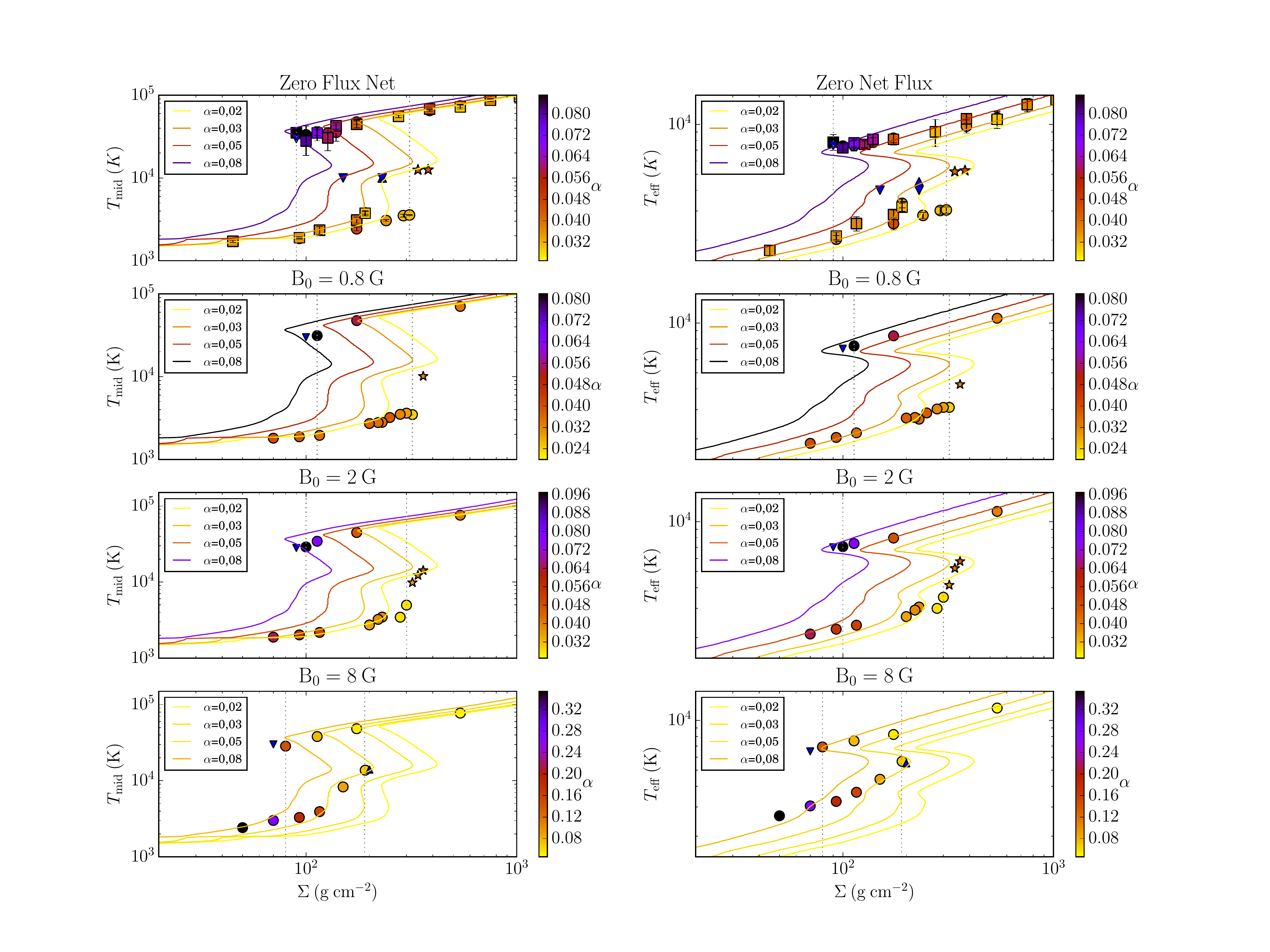}
\caption{Thermal equilibrium curves in the $\Sigma$-$T_\mathrm{mid}$($-T_\mathrm{eff}$) plane for different magnetic field configurations. From top to bottom, the first panel shows the result of S18 (circles) with outflow conditions with new additional simulations using a zero net magnetic field configuration, the second panel uses $\mathrm{B_{z0}=0.8\:G}$, the third panel $\mathrm{B_{z0}=2\:G,}$ and the last panel $\mathrm{B_{z0}=8\:G}$. The colors of the dot correspond to the value of $\alpha$. Triangles show the instability points where the patch of disk undergoes critical cooling (downward-pointing triangle) or heating (upward). Stars indicate the simulations from the middle branch that did not converge. Gray vertical lines are plotted at the end of the hot and cold branch to facilitate comparisons between the panels. Finally, the colored S-curves in solid lines shows the DIM model with convection using different values for $\alpha$.}
\label{S8G}
\end{figure*}

We plot in the left (right) panels of Figure \ref{S8G} the thermal equilibrium curves in the $T_\mathrm{mid}$($T_\mathrm{eff}$) versus $\Sigma$ plane, also known as S-curves, for several net magnetic fields. We emphasize that for a given S-curve, $B_{z_{0}}$ is fixed, but not $\beta$. The S-curve in ZNF is taken from S18 with additional simulations with outflow $z$-boundary conditions on the cold branch to compare this with the net flux case. These additional simulations show better agreement than S18 with \cite{Hirose2014} concerning the cold branch.

Returning to the net flux case, we see that similarly to ZNF simulations, there are always two branches of stability: the hot branch, and the cold branch. However, as observed in \cite{hirose2015} and S18, there is also a middle branch, which is a continuation at higher temperature of the cold branch; these points are indicated as stars in Figure (\ref{S8G}). There is no obvious middle branch in the case of $B_{z_{0}}$=8 G since the temperature on the cold branch gradually increases from 3005 K to 13751 K. For $B_{z_{0}}$=0.8, 2 G and ZNF we note that the middle branch does not converge with the box size. The equilibrium temperature strongly depends on the box size. We do not know the origin of this dependence. Given the uncertainties regarding the stability of these points, we did not include these simulations in any other figure (see also \cite{hirose2015}).

The extrema of the S-curves, defined by $\Sigma_\mathrm{crit}^+/\Sigma_\mathrm{crit}^-$ and $T_\mathrm{eff\:crit}^+/T_\mathrm{eff\:crit}^-$ , are of special importance as they determine the shape of the light curves and in particular the presence of reflares (\citealt{lasota2001}; \citealt{coleman2016}). For $B_{z_{0}}$<8 G, the S-curves are not very different from the ZNF case. We find $\Sigma_\mathrm{crit}^-\approx110\:\mathrm{g\:cm^{-2}}$ , which is compatible with S18 given the uncertainty on the stability limit.  However, for $B_{z_{0}}$=8 G, we have $\Sigma_\mathrm{crit}^-$=80 $\mathrm{g\:cm^{-2}}$ , which extends the hot branch to lower densities. The same result applies to the beginning of the middle branch. However, the conclusions concerning $\Sigma_\mathrm{crit}^+$ are more difficult to draw from our simulations as we see a box-dependent behavior of the middle branch, as reported above. In the DIM, S-curves with higher $\alpha$ are shifted toward lower densities. Since increasing the magnetization gave higher values of $\alpha$, our results are consistent with the DIM. 

The S-curve with $B_{z_{0}}$=8 G has another notable feature compared to the lower net magnetic field S-curves and S18: here alone, the maximum value of $\alpha$ is obtained on the cold branch. We postpone the discussion of this point to \S\ref{sec:discussion}. For $\Sigma<116\:\mathrm{g\:cm^{-2}}$ , magnetization starts to be important on the cold branch with $B_{z_{0}}$=2 G and $\alpha$ is greater than the ZNF case. Otherwise, in the remaining S-curve and for weaker magnetic field strengths, the values of $\alpha$ are similar to the ZNF case.

Finally, we note that on the hot branch our simulations are well fit by the S-curves from the DIM. However, simulations with the strongest magnetization, which are located on the cold branches  of cases $B_{z_{0}}$=2 and 8 G, clearly do not fit the equilibrium curves from the DIM. As the disk instability model is a purely hydrodynamic code, this is not surprising. 

\subsection{Vertical and radial transport of angular momentum\label{sec:accretion}}
In a disk, the removal of angular momentum from turbulent MRI and surface torque leads to an inward flow of matter along the radial direction. We define the mass accretion rate in the disk $\dot{M}$ as \begin{equation*}
\dot{M}\equiv -2\pi R_0\Sigma\braket{u_R}_{\rho.}
\end{equation*}
From the conservation of angular momentum, we find that
\begin{equation*}
\dot{M}=\frac{4\pi}{\Omega_0}\left(\frac{1}{R_0}\partial_{R_0}(\Sigma R_0^2\alpha\braket{c_s^2}_{\rho})+\frac{2q}{\beta}\braket{P_\mathrm{thermal\:mid}}R_0\right)
\end{equation*}
Following \cite{fromang2013}, we decompose $\dot{M}$ into a radial component, $\dot{M}_{R\phi}$, and a vertical component, $\dot{M}_{z\phi}$, of angular momentum transport,
\begin{gather}
\dot{M}_{R\phi}=\frac{4\pi}{\Omega_0}\frac{1}{R_0}\partial_{R_0}(\Sigma R_0^2\alpha\braket{c_s^2}_{\rho}), \nonumber\\
\dot{M}_{z\phi}=\frac{8\pi }{\beta}q\braket{P_\mathrm{thermal\:mid}}\frac{R_0}{\Omega_0}. \label{eq:mzphi}
\end{gather}
Because the curvature of the shearing box is absent, the term $\dot{M}_{R\phi}$ measured in our simulation is zero. However, by making a few assumptions on the radial structure of the disk, we can have an estimate of $\dot{M}_{R\phi}$. We assumed the following scalings: $\Sigma\approx R^p$ and $\braket{c_s^2}_{\rho}\approx R^s$. Equation (\ref{eq:mzphi}) then becomes
\begin{equation*}
\dot{M}_{R\phi}=\frac{p+s+2}{\Omega_0}4\pi\Sigma\alpha\braket{c_s^2}_{\rho.}\\
\end{equation*}
Classical viscous disk theory predicts $\Sigma\propto R^{-3/4}$ and $T_\mathrm{eff}\propto R^{-3/4}$ \citep{Frank}. This means that $(p+s+2)$ is not far from unity. The ratio of the radial to vertical accretion rate is then
\begin{equation*}
\frac{\dot{M}_{R\phi}}{\dot{M}_{z\phi}}=(p+s+2)\frac{\beta}{2q}\frac{\Sigma\braket{c_s^2}_{\rho}}{\braket{P_\mathrm{thermal\:mid}}}\frac{\alpha}{R_0.}
\end{equation*}
From our simulations, we find that $\Sigma\braket{c_s^2}_{\rho}/\braket{P_\mathrm{thermal\:mid}}\approx C(\beta,H)\times H,$ where the factor $C(\beta,H)$ extends from 7.4 to 1.6 in our simulations. This means that\begin{equation*}
\frac{\dot{M}_{R\phi}}{\dot{M}_{z\phi}}\approx C\alpha\times\frac{\beta}{2q}\left(\frac{H}{R_0}\right)
\end{equation*}
We can use the empirical relations from Figures \ref{alpha_beta} and \ref{q_over_beta_vs_beta}, and we find 
\begin{equation} \label{eq:empirical}
\frac{\dot{M}_{R\phi}}{\dot{M}_{z\phi}}\approx \begin{cases}0.19C\times\beta^{0.44}\left(\frac{H}{R_0}\right), & \mbox{if } \beta<10^4 \\ 0.30C\times\beta^{0.4}\left(\frac{H}{R_0}\right), & \mbox{if } \beta>10^4 \end{cases} 
.\end{equation}

\begin{figure}[h!]
\centering
\includegraphics[width=100mm,height=90mm]{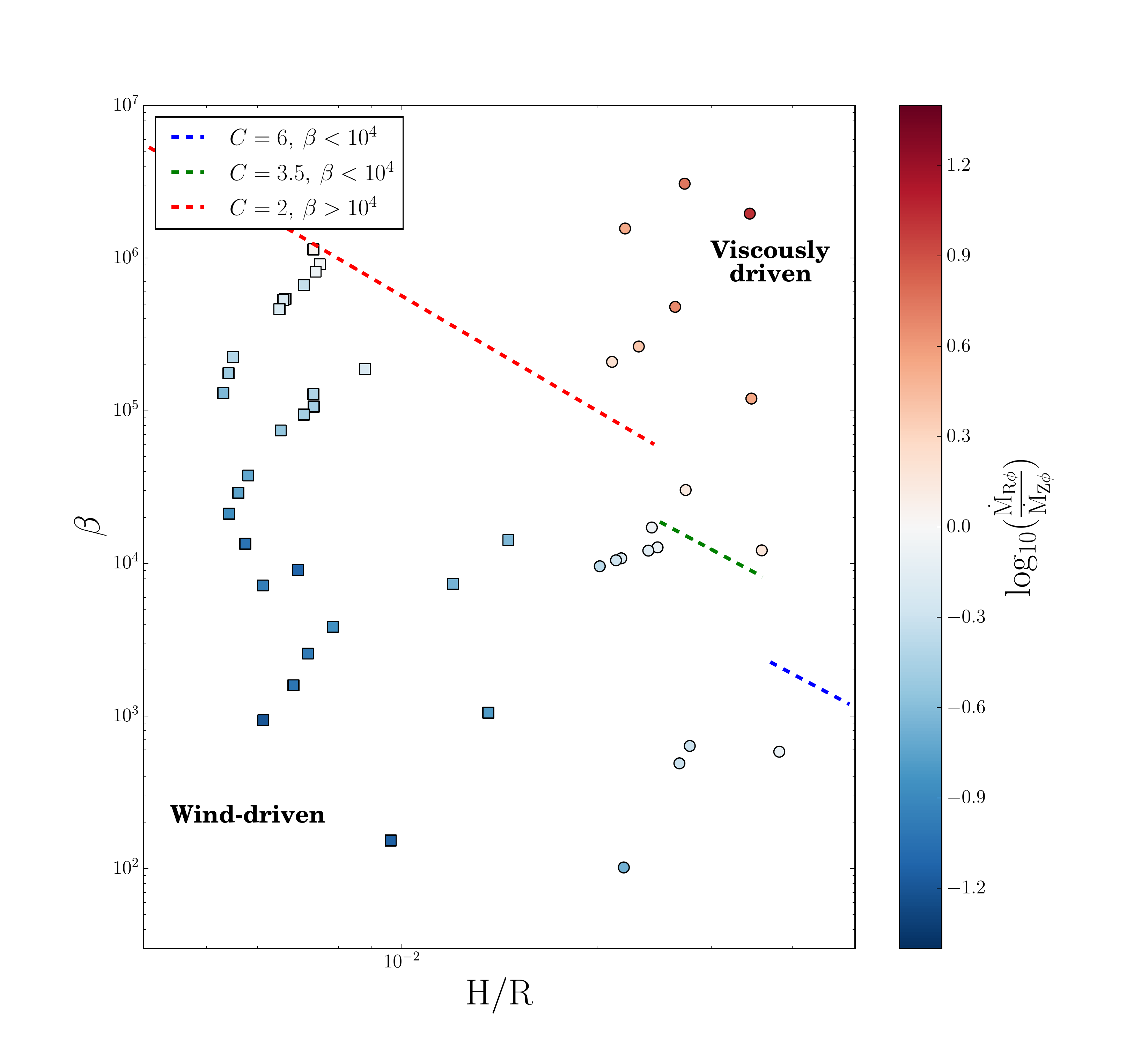}
\caption{Ratio of the accretion rate due to radial transport of angular momentum $\dot{M}_{R\phi}$ to the accretion rate due to vertical transport of angular momentum $\dot{M}_{z\phi}$ as a function of $\beta$ and $\mathrm{H/R}$. The color shows the value of $\log(\dot{M}_{R\phi}/\dot{M}_{z\phi})$. Squares indicates cold branch simulations and circles hot branch simulations. Dashed lines show the limit $\dot{M}_{R\phi}/\dot{M}_{z\phi}=1$ from the empirical formulas given by Eq. (\ref{eq:empirical}).}
\label{Mr_Mz}
\end{figure}

The values of $\dot{M}_{R\phi}$ and $\dot{M}_{z\phi}$ can be found in Table \ref{tab}. We plot in Figure \ref{Mr_Mz} the ratio of the radial to vertical accretion rate as a function of $\beta$ and $H/R$. Dashed lines show the limit $\dot{M}_{R\phi}/\dot{M}_{z\phi}=1$ from the empirical formulas given by Eq. (\ref{eq:empirical}). Simulations from the hot and cold branch are represented by circles and squares, respectively. Figure 5 shows that the cold branch is entirely dominated by accretion that is due to vertical transport of angular momentum. Only simulations from the denser and hotter part of the hot branch are dominated by viscous radial transport, and even in this case, vertical transport remains non-negligible. This shows that the vision of a purely viscous $\alpha$-disk is inappropriate to describe an DNe disk if a large-scale poloidal field is present. In such a case, the role of the wind torque in removing angular momentum must be taken into account when computing the disk evolution. This can be done using the fits we provide for $q_\mathrm{surface}$ (see Figure \ref{q_over_beta_vs_beta}). Additionally, the limit $\dot{M}_{R\phi}/\dot{M}_{z\phi}=1$ is well reproduced by empirical formulas from Eq. (\ref{eq:empirical}), providing a good estimate of the constant C. As a comparison, in the global simulations from \cite{zhu2018}, where $H/R=0.1$ and $\beta\approx10^3$, the transport is dominated by the viscous torque with a contribution of only $5\%$ from the wind torque. Using our empirical formula with $C=\sqrt{2\pi}$, we find that the wind torque accounts for $\approx20\%$ of the total torque. Again, our simple shearing-box estimates of the wind torque are on the same order of magnitude as global simulations.

\subsection{Stability criterion in a disk-wind model\label{sec:DIM_wind}}
The essence of the DIM is the thermal instability around ionization of hydrogen. In the DIM, the effective temperature is directly related to the mass accretion rate in the disk by 
\begin{equation}
\sigma T_\mathrm{eff}^4=\frac{3GM\dot{M}}{4\pi R_0^3}\left(1-\left(\frac{R_{in}}{R_0}\right)^{1/2}\right)
.\end{equation}
In a stationary system, the mass accretion rate is imposed by the mass transfer rate from the companion. Hence, in a range of $\dot{M,}$  part of the disk is ionizing hydrogen and thus unstable; this corresponds to DNe. If the accretion rate is high enough for the annulus at the external radius to be fully ionized, then the disk is stable; this a novae-like system. This simple argument provides an observational test of the DIM. \cite{dubus2018} systematically examined the relation between $\dot{M}$ and the stability of the system for $\approx$100 CVs and found that predictions based on the DIM are robust. 

We find that $T_\mathrm{eff\:crit^-}$ does not change much between ZNF and net flux simulations. It is determined by the ionization temperature of hydrogen and is $\approx7000\:K$. However, when we include a contribution from the wind in the conservation of energy and angular momentum, we find that the classical relation between $\dot{M}$ and $T_\mathrm{eff}$ in the disk changes to
\begin{align} \label{eq:heating}
\sigma T_\mathrm{eff}^4&=\frac{3GM\dot{M}}{4\pi R_0^3}\left(1-\left(\frac{R_{in}}{R_0}\right)^{1/2}-\frac{1}{2\dot{M}R_0^{1/2}}\int_{R_{in}}^{R_0}\dot{M}_{z\phi}r^{-1/2}dr\right)\\ \label{eq:heatingb}
&=\frac{3GM}{8\pi R_0^{7/2}}\int^{R_0}_{R_\mathrm{in}}\dot{M}_{R\phi}r^{-1/2}dr.
\end{align}
We can rewrite this equation as
\begin{equation}
\sigma T_\mathrm{eff}^4=\frac{3GM\dot{\mathcal{M}}_\mathrm{vis}}{4\pi R_0^3}\left(1-\left(\frac{R_{in}}{R_0}\right)^{1/2}\right)
,\end{equation}
where $\dot{\mathcal{M}}_\mathrm{vis}$ corresponds to the viscous mass transfer rate, which a purely viscous disk would have to match the dissipation rate of a turbulent, wind-driven disk.

In the presence or absence of wind-driven accretion, heating is only due to viscous accretion. This implies that $W_{R\phi}$ must be positive since it is of the same sign as the heating, which, by definition, is positive (see \cite{balbus98}). With the assumption that $W_{R\phi}(R_\mathrm{in})$=0 in a purely viscous disk,  $\dot{M}_\mathrm{R\phi}$ needs to be constant and positive to ensure stationarity. This is no longer true when angular momentum is partly extracted by a wind. We see from Eq. (\ref{eq:heatingb}) that in a stationary disk, $\dot{M}_{R\phi}$ does not have to be positive at all radii or constant in order to ensure positive heating. This breaks down the classical relation between the viscous mass accretion rate and effective temperature \citep{pringle1981} and opens up the possibility for more complex disk structures in which the disk can be viscously excreting and still be in steady state.

Equation (\ref{eq:heating}) implies that a system with a high $\dot{M}$ could be unstable if there is a significant contribution to the angular momentum transport from the wind torque. The term in parentheses on the right-hand side of Eq. (\ref{eq:heating}) is positive and smaller than unity. Hence, the disk heating is weaker than would be deduced from a viscous model with the same total $\dot{M}$. The disk temperature can become very low if most of the transport is due to the wind (with no associated heating), so that even a disk with a high total $\dot{M}$ could reside on the cold branch.  \cite{dubus2018} computed the mass accretion rates from optical luminosities, hence they measured $\dot{\mathcal{M}}_\mathrm{vis}$ , that is,  a measure of the viscous dissipation rate, but not of the total $\dot{M}$ or of $\dot{M}_{R\phi}$. The separation they found between stable and unstable systems should still hold since the disk instability criterion is fundamentally based on temperature, hence on $\dot{\mathcal{M}}_\mathrm{vis}$. However, there could be a large discrepancy between $\dot{\mathcal{M}}_\mathrm{vis}$ and the total $\dot{M}$ , which could be tested if a direct estimate of the mass transfer from the companion $\dot{M}$ were to be available.

\section{Resistive MHD runs\label{sec:resistive}}
Ideal MHD is a very poor approximation on the cold branch. Previous results from S18 showed that in a zero net flux configuration, a threshold in temperature exists below which MRI can no longer sustain turbulence. This transition occurs when the magnetic Reynolds number $R_m$ is lower than $\approx 5\times 10^3$ with
\begin{equation*}
R_m=\frac{c_sh}{\eta}.
\end{equation*}

In the ZNF case, the flow has to sustain the field for the MRI to operate through a dynamo feedback loop. This positive feedback is not required in the net flux case since the field is provided by the environment, such that turbulent motions can be excited at lower $R_\mathrm{m}$. Therefore the presence of a net $\braket{B_z}$ reduces the threshold below which linear MRI cannot grow and offers the possibility for MRI to survive down to lower ionization fractions. This behavior was observed in \cite{fleming2000}, who described a cyclic revival of MRI down to $R_m=50$ compared to the case of ZNF, where MRI turbulence disappears for $R_\mathrm{m}<10^4$. However, \cite{fleming2000} considered a uniform fixed vertical profile of resistivity with an isothermal equation of state, which can lead to an artificial quasi-steady state. We wish to  determine whether by relaxing this constraint, the patch of disk can keep turbulence active to lower ionization fractions than in ZNF.

Here, resistivity was computed as in \cite{blaes1994}:
\begin{equation*}       
\eta=230\left(\frac{n_n}{n_e}\right)T^{1/2}\:\mathrm{cm^2\:s^{-1}}
\end{equation*}
with the assumption that this is  dominated by electron-neutral collision. The values of $\eta$ were pre-computed with the equation of state (Saha equilibrium) and saved in a table.

We restarted from previous ideal MHD simulations on the cold branch and added resistivity. We imposed a floor of 50 on the magnetic Reynolds number in order to maintain a reasonably large time step. 

\subsection{Impact of $B_z$ on the resistive cold branch}
$R_\mathrm{m,\:turb}$ is the critical Reynolds number below which no turbulence is seen in our simulation. It is computed as the midplane value of $R_m$ at the moment where the Maxwell stress ultimately decays to zero. 

\begin{figure}[h!]
\centering
\includegraphics[width=100mm,height=70mm]{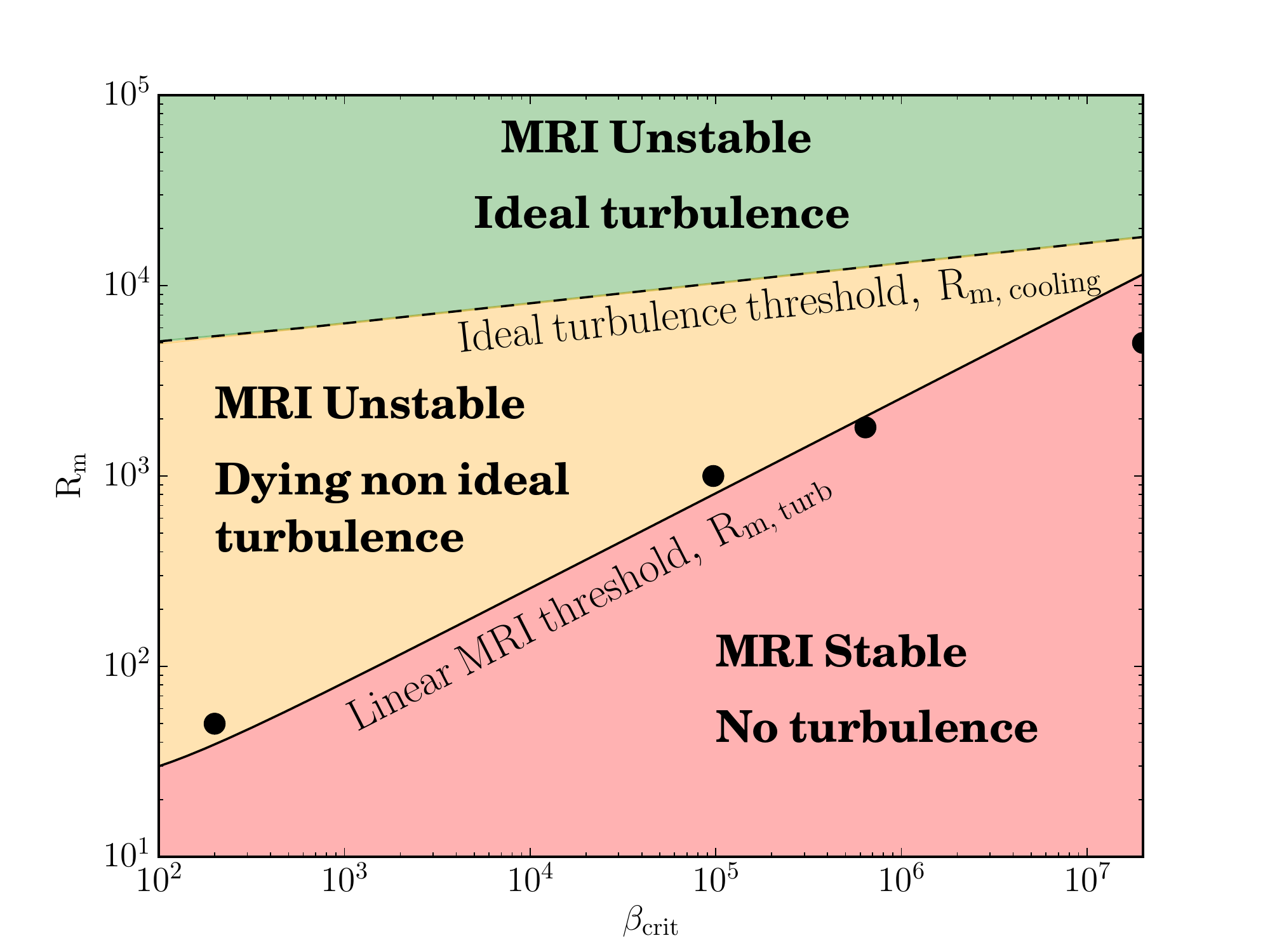}
\caption{Dots show the magnetic Reynolds number $R_\mathrm{m,\:turb}$ for which turbulence ceases in our simulation as a function of $\beta_\mathrm{turb}$. The oblique solid line is drawn from Eq. (36), which gives $R_\mathrm{m,\:linear}$, the linear criterion for stability of resistive MRI. The oblique dashed line gives the $R_\mathrm{m,\:cooling}$ below which turbulence is affected by resistivity and will ultimately be suppressed.}
\label{Rm}
\end{figure}

Following \cite{jin1996}, we find that the linear criterion for stability of resistive MRI in the case of a vertical field and no stratification is 
\begin{equation*}
R_\mathrm{m,\:linear}=\frac{\beta}{\sqrt{\frac{3}{2\pi^2}\beta-4.}}
\end{equation*}
We plot in Figure \ref{Rm} $R_\mathrm{m,\:turb}$ as a function of $\beta$ together with its analytical estimate from linear theory. As expected, the threshold for turbulence decreases with stronger magnetization. Our results agree with the linear criterion plotted as an oblique solid black line on Figure \ref{Rm}.

Nonetheless, we observe that turbulence is weaker than the pure ideal case even for $R_\mathrm{m}>R_\mathrm{m,\:turb}$. In $407F\_B0.8G,$ for example, the Reynolds magnetic number is larger than $\mathrm{R_{m,\:turb}}$ for $\approx100$ orbits after we added resistivity. The turbulence is affected by resistivity, however; $\alpha$ slowly decreases, causing the disk to cool until it reaches the point where $R_\mathrm{m}<R_\mathrm{m,\:turb}$ and turbulence ceases. This shows that a region exists where MRI is still linearly unstable but turbulence irremediably ceases. The reason is that as T decreases, $\alpha$ decreases faster than $Q^-$, leading to a thermally unstable situation. The timescale on which this occurs is inversely proportional to the initial $R_\mathrm{m}$ from which we started resistivity. For $R_\mathrm{m}\approx 1.5\times10^4$ and $5\times10^3$ , we have a time decay for the Maxwell stress of $\approx100$ and $20$ orbits, respectively. 

Based on these observations, we define another parameter, $R_\mathrm{m,\:cooling}$, below which turbulence is still active but not self-sustaining. $R_\mathrm{m,\:cooling}$ does not change much between ZNF where $R_\mathrm{m,\:cooling}=18000$ and net flux simulations with $B_{z_{0}}$=8 G where $R_\mathrm{m,\:cooling}=6000$. We summarize these results in Figure \ref{Rm}, where the dashed black line shows $\mathrm{R_{m,\:cooling}}$ from our simulations. Three distinct zones are evident: for $R_\mathrm{m}>R_\mathrm{m,\:cooling}$ , turbulence is similar to the ideal case; for $R_\mathrm{m,\:cooling}>R_\mathrm{m}>R_\mathrm{m,\:turb}$ , turbulence ceases and the disk cools until it reaches the zone $R_\mathrm{m}<R_\mathrm{m,\:turb}$ , where MRI is stable.

These results show that ideal turbulence can be maintained to slightly lower ionization levels in the presence of a large-scale magnetic field. This effect is too subtle for $B_{z_{0}}$=0.8 and 2 G to maintain the cold branch, however. Only for $B_{z_{0}}$=8 G is the cold branch maintained down to $\Sigma=70\:\mathrm{g\:cm^{-2}}$. Surely, the fact that $R_\mathrm{m,\:cooling}=6000$ helps in sustaining the cold branch to such low surface densities, but the main actor is that $\alpha$ takes high values, up to 0.275, because of the low $\beta$. This enhanced heating ensures a level of thermal ionization that is higher than in simulations with weaker magnetic fields. In this case, two effects act together in keeping the MRI unstable: first and foremost, the enhanced heating at high magnetization, and secondarily, the effect of net magnetic flux on resistive MRI.

\subsection{Transport of angular momentum from the upper ionized layers}
It is probable that even with a net magnetic field, there are regions of the disk that are too cold to sustain MRI turbulence. In this last section, we therefore study the effect of X-ray irradiation on the regions where turbulence has ceased. S18 showed that X-ray ionization from the central white dwarf (WD) cannot increase the ionization fraction in the midplane above the threshold required for MRI to be unstable. This is consistent with the picture described in \cite{gammie1996}, where an X-ray illuminated accretion disk of DNe in quiescence was organized into layers with a so-called "dead zone" in the central region. However, S18 neglected the active zones in the upper layers where the ionization fraction suffices to ensure coupling between the field and the plasma. As in protoplanetary disks, this could ensure transport of angular momentum from the wind torque (\citealt{bai2013b}, \citealt{bethune2017}). The ensuing supersonic accretion of the upper layers of the disk could then lead to the required accretion rates. Following the procedure developed in \S\ref{sec:accretion}, we can estimate that for $\alpha\approx0.035$ and $\Sigma\approx100\:\mathrm{g\:cm^{-2}}$ , the typical turbulent accretion rate is $\mathrm{\dot{M}_{R\phi}}\approx5\times10^{14}\mathrm{\:g\:s^{-1}}$. We wish to estimate the ionization fraction in the upper layers and determine whether we can obtain such an accretion rate.

Following \citet{glassgold1997,1997ApJ...485..920G}, we computed the ionization rate $\zeta$ assuming a photon flux 
\begin{equation*}
f_0 = {\cal C}\frac{L_{XR}}{4\pi R_0^2kT_{XR,}}
\end{equation*}
where ${\cal C}$ takes into account the irradiation geometry and the albedo $A$ of the disk. S18 estimated that $C=10^{-2}$. From \cite{byckling2010}, we adopted a maximum X-ray flux $L_{XR}$ of photons emitted from the white dwarf of $\mathrm{10^{31}\:erg\:cm^{-2}}$ with a bremsstrahlung spectrum of temperature $\mathrm{T_{XR}}$=10 keV. We adopted a characteristic surface density of 100 $\mathrm{g\:cm^{-2}}$.

We find that $\approx1\%$ of the mass is ionized in the upper active layers. From now on, we assume that only the upper ionized layers of the disk accrete at a constant velocity $\mathrm{v_R}$. The typical accretion rate on the cold branch being $\mathrm{\dot{M}}\approx2\pi R\Sigma\braket{v_R}_{\rho}\approx 5\times10^{14}\:\mathrm{g\:s^{-1}}$ , this leads from a surface density of 100 $\mathrm{g\:cm^{-2}}$ to $\mathrm{v_R}\approx5\times10^3\:\mathrm{cm\:s^{-1}}$. From the equation of angular momentum we can obtain the following relation between $q$ and the accretion speed:
\begin{equation*}
v_R=\frac{2qB_z^2}{3\Omega\Sigma f_\mathrm{ioniz,}}
\end{equation*}
which shows that the minimum wind torque that could explain the accretion rate on the cold branch is $q_\mathrm{min}\approx10$. From Figure \ref{alpha_beta}, we find for $\beta<10^4$ that $q_\mathrm{surface}$ takes a constant value of $\approx40,$ which is above $q_\mathrm{min}$. Adding the caveat that $q_\mathrm{surface}$ is only an estimate from a shearing box in ideal MHD, we propose sonic accretion in the upper layers as a reasonable option to explain accretion in resistive regions of DNe.

\section{Discussion \label{sec:discussion}}
S18 have confirmed that MRI leads to values of $\alpha\approx0.1$ near the tip of the hot branch when coupled non-linearly with convection. When we add a net magnetic flux, we find that convection also enhances $\alpha$ for magnetizations up to $\beta\approx10^4$ , showing that convective MRI is a reliable mechanism for enhancing classical MRI transport. The enhancement of $\alpha$ is localized at the tip of the hot branch and cannot explain values of 0.1 in the high-density part of the hot branch, however. This could be obtained if $\beta$ were high throughout the whole hot branch. However, a strong vertical magnetic field also leads to high values of $\alpha$ on the cold branch. Obviously, the large magnetic field is surely not constant throughout a hysteresis cycle of DNe. Then, it is possible that the advection of the magnetic field leads to a higher magnetization during the eruptive phase than in quiescence. Unfortunately, advection of a large-scale magnetic field is a global phenomenon that cannot be modeled in the shearing-box framework (see \citealt{guilet2012}, \citealt{guilet2014} for recents developments on that topic).

We investigated the problem of angular momentum transport on the cold branch and found that resistive MRI does not naturally lead to values of $\alpha\approx0.01,$ as required by the DIM models. S18 have indeed shown that when resistivity is included in a ZNF configuration, the cold branch becomes MRI stable and the transport is quenched. The same conclusion applies for this paper in the case of $B_\mathrm{z}$=0.8 and 2 G. For $B_\mathrm{z}$=8 G, the disk remains turbulent on most of the cold branch, but exhibits values of $\alpha\gtrsim0.1$. 

Some net magnetic flux is unavoidable in the disks of DNe since the white dwarf or companion are likely to have a non-zero dipolar field. If this leads to low values of $\beta,$ then the mass accretion is mostly due to the torque from the magnetic wind that appears. This will have an effect on the evolution of the disk throughout an outburst cycle. Wind transport will also change the link between light-curve timescales and $\alpha$. In a purely turbulent case, the light-curve timescales are related to the thermal ($t_{\rm therm}\propto 1/\alpha\Omega$) and viscous ($t_{\rm vis}\sim R^2/\nu$) timescales of the disk. For instance, the propagation time of a cooling front is $\sim (t_{\rm therm}t_{\rm vis})^{1/2}$ (\citealt{menou1999}; \citealt{kotko2012}).  When a wind is present, however, the observed timescales will become a diffusion or advection time related to $\alpha,$ but also to $q$. This will have to be borne in mind in future determinations of $\alpha$ from observations.

To build such a disk-wind model, we provided prescriptions for $\alpha$ and the wind torque $q$ estimated from our shearing-box simulations. We compared our results with local simulations from \cite{bai2013a}, \cite{fromang2013}, and \cite{simon2013a} as well as global simulations from \cite{zhu2018}, and we found that our estimates of the wind torque and the mass accretion rates agree with those in the literature. However, an uncertainty remains concerning the dependence of $q$ on $H/R$. This problem needs to be addressed with global simulations.

On a more observational side, we know that there are winds in the eruptive state of DNe that are emitted by the disk \citep{cordova1982}, which may be hydromagnetically launched \citep{cannizzo1988b}. However, only one potential indirect detection of wind in quiescence has been reported by \cite{perna2003}. The detection is expected to be more difficult than for a disk in eruption (\citealt{drew1990}; \citealt{perna2003}), and it should not be concluded from the lack of observational evidence that winds are absent in quiescence.

Mass accretion rates computed through X-ray emission allow access to the total mass accretion rates. Several sources indicate that the X-ray flux in quiescence is higher than what can be explained by the classical DIM (\citealt{collins2010}, \citealt{mukai2017}, \citealt{wheatley2003}). One explanation is that the disk may be truncated by the magnetic field of the WD (\cite{wheatleywest2003}; \citealt{balman2012}). However, this hypothesis is still debated, and there is also strong evidence of an accretion disk that even reaches the WD (\citealt{mukai2017}; \citealt{ishida2009}). In view of our results, we suggest that accretion in quiescence is mainly driven by the wind torque. This would reconcile the high X-ray flux coming from the flow of matter $\dot{M}$ onto the white dwarf with the low values of $\dot{M}_{R\phi}$ required by the DIM: in such a case the dissipation in the disk is related to $\dot{M}_{R\phi}$ and not to $\dot{M}$, leading to a darker disk than a turbulence-driven disk with the same $\dot{M}$.

We also propose that a plausible solution for the colder regions of the disk is accretion that is not driven by turbulence, but by removal of angular momentum by the wind torque in the ionized upper layers, much as in protoplanetary disks. We used observational constraints of the X-ray flux in quiescence emitted from the central white dwarf to compute the ionization level in the upper layers. We estimate that it might be sufficient to explain the $\alpha$ that are observed. A layered accretion is likely to have an effect on how outbursts are triggered in the disk, notably on the conditions for which the heating front propagates outward ( "inside-out outbursts") or inward ("outside-in"). There is observational evidence for both in dwarf novae. In this context, the absence of heating from the wind is a major problem. This may prevent the disk from reaching high enough temperatures to ionize H and trigger the thermal-viscous instability. The trigger may then be due to a decrease in the wind torque as matter accumulates in quiescence, a profound modification to the DIM. Moreover, the lack of heating may not enable maintenance of the vertical structure, and some base level of heating is required in the poorly ionized parts of the disk.

\section{Conclusion}
We have studied the impact of a large net vertical magnetic field on the properties of transport and on the thermal equilibrium curves of DNe. We considered a large-scale magnetic field as could arise from the dipolar component of the primary or the companion, which is constant for a given S-curve. This configuration naturally leads to lower values of $\beta$ on the cold branch than on the hot branch. For $B_z<2G$, the S-curves, in particular the critical points, are quasi-identical to the ZNF case. For $B_z\gtrsim2$ G, the S-curves are shifted toward low density, and we found higher values of $\alpha$ on the cold branch than on the hot branch.

We investigated the problem of the transport of angular momentum on the cold resistive branch. We find that for $B_\mathrm{z}$=0.8 and 2 G, MRI turbulence is suppressed on most of the cold branch. However, for $B_\mathrm{z}$=8 G, the turbulence is sustained to lower surface densities, but the cold branch exhibits values of $\alpha>0.1$. 

We showed that for $R=1.315\times10^{10}$ cm, the entire cold branch is wind driven, even for local magnetic fields that are relatively weak. It is therefore essential to include wind-driven transport in the DIM and to review the effect on our understanding of outburst cycles.

Outflows provide an alternative route for extracting angular momentum that has a huge impact on the budget of angular momentum, but does not heat the disk. This means that DNe may be accreting much more than what we infer from the luminosity of the disk, and the high X-ray flux observed in quiescence may support this theory. All in all, we arrive at a point where observations and theory need to work hand in hand again if we wish to proceed in our understanding of accretion in DNe. 

\begin{acknowledgement}
NS, GL, and GD thank Omer Blaes for fruitful discussions. We thank the anonymous referee for insightful comments. NS acknowledges financial support from the pole PAGE of the Universit\'e Grenoble Alpes. GL, GD, and MF are grateful to the participants of the KITP 2017 program on {\em Confronting MHD Theories of Accretion Disks with Observations} for the many useful conversations pertaining to this work (and thus this research was supported in part by the National Science Foundation under Grant No. NSF PHY-1125915). This work was granted access to the HPC resources of IDRIS under the allocation A0020402231 made by GENCI (Grand Equipement National de Calcul Intensif). Some of the computations presented in this paper were performed using the Froggy platform of the CIMENT infrastructure (https://ciment.ujf-grenoble.fr), which is supported by the Rhône-Alpes region (GRANT CPER07\_13 CIRA), the OSUG@2020 labex (reference ANR10 LABX56) and the Equip@Meso project (reference ANR-10-EQPX-29-01) of the programme Investissements d'Avenir supervised by the Agence Nationale pour la Recherche. M. Flock has received funding from the European Research Council (ERC) under the European Union's Horizon 2020 research and innovation programme (grant agreement No 757957).
\end{acknowledgement}

\bibliographystyle{aa}
\bibliography{biblio}

\end{document}